\documentclass{aa}  

\usepackage{graphicx}
\usepackage{txfonts}

\newcommand{\kmos}{\mbox{\,km\,s$^{-1}$}}
\newcommand{\mos} {\mbox{\,m\,s$^{-1}$}}

\begin{document}     
   
   \title{Spectroscopic properties of 
   \\ a two-dimensional time-dependent Cepheid model 
   \\ I. Description and validation of the model}

  \subtitle{}
  
   \author{V. Vasilyev\inst{1,2},
           H.-G. Ludwig\inst{1},
           B. Freytag\inst{3},
           B. Lemasle\inst{4},
            \and M. Marconi\inst{5}
          }
\institute{%
Zentrum f\"ur Astronomie der Universit\"at Heidelberg, Landessternwarte,
K\"onigstuhl 12, D-69117 Heidelberg, Germany \label{1}
\and
Max-Planck-Institut f\"ur Astronomie(MPIA), K\"onigstuhl 17,
D-69117 Heidelberg, Germany, 
\email{vasilyev@mpia-hd.mpg.de}\label{2}
\and
Department of Physics and Astronomy at Uppsala University, 
Regementsv\"agen 1, Box 516, SE-75120 Uppsala, Sweden\label{3}
\and
Zentrum f\"ur Astronomie der Universit\"at Heidelberg, Astronomisches Recheninstitut, Mönchhofstr. 12-14, D-69120, Heidelberg, Germany\label{4}
\and
INAF -Osservatorio Astronomico di Capodimonte, Via Moiariello 16, I-80131
Napoli, Italy\label{5}
    }
   \date{Received ; accepted }
  
\abstract{Standard spectroscopic analyses of Cepheid variables are based on hydrostatic
  one-dimensional model atmospheres, with convection treated using
  various formulations of mixing-length theory.}
{This paper aims to carry out an investigation of the validity of the quasi-static approximation in the context of
  pulsating stars. We check the adequacy of a two-dimensional
  time-dependent model of a Cepheid-like variable with focus on its
  spectroscopic properties. }
{With the radiation-hydrodynamics code CO5BOLD, we construct a two-dimensional
time-dependent envelope model of a Cepheid with  $T_\mathrm{eff}= 5600$ K,
$\log g=2.0$, solar metallicity, and a 2.8-day pulsation period. Subsequently,
we perform extensive spectral syntheses of a set of artificial iron lines in local
thermodynamic equilibrium. The set of lines allows us to
systematically study effects of line strength, ionization stage, and excitation
potential.}
{We evaluate the microturbulent velocity, line asymmetry, projection factor,
  and Doppler shifts.  The microturbulent velocity, averaged over all lines,
  depends on the pulsational phase and varies between 1.5 and 2.7\kmos. The
  derived projection factor lies between 1.23 and 1.27, which agrees with
  observational results.  The mean Doppler shift is non-zero and negative,
  -1\kmos, after averaging over several full periods and lines.  This residual
  line-of-sight velocity (related to the ``K-term'') is primarily caused by
  horizontal inhomogeneities, and consequently we interpret it as the familiar
  convective blueshift ubiquitously present in non-pulsating late-type
  stars. Limited statistics prevent firm conclusions on the line
  asymmetries.}
{Our two-dimensional model provides a reasonably accurate representation of
  the spectroscopic properties of a short-period Cepheid-like variable
  star. Some properties are primarily controlled by convective inhomogeneities
  rather than by the Cepheid-defining pulsations. Extended multi-dimensional
  modelling offers new insight into the nature of pulsating stars.}

   \keywords{methods: numerical -- radiative transfer -- convection -- stars: atmospheres  -- stars: oscillations -- stars: variables: Cepheids 
   }
      \titlerunning{Spectroscopic properties of a 2D Cepheid model. I}
   \authorrunning{V.~Vasilyev et al.}  
  \maketitle

%

\section{Introduction}

Cepheid variable stars play an important role in astronomy.  In cosmological
applications they are standard candles to measure Galactic and extragalactic
distances with unprecedented precision \citep{2016ApJ...826...56R} due to
a remarkable relation between the pulsation period and their intrinsic
luminosity \citep[the Leavitt
  law:][]{1908AnHar..60...87L,1912HarCi.173....1L}.  \text{The period-luminosity relation}
can provide
measurements of the distance up to approximately 30\,Mpc. 
Despite the overall successful application of the period-luminosity relation, its
universality has been one of the most debated issues concerning Cepheids in
the last few decades \citep[see e.g.][and references
therein]{2005ApJ...632..590M, 2008A&A...488..731R, 
2009MmSAI..80..141M, 2010ApJ...713..615M, 
2010ApJ...715..277B, 2017arXiv170510855W}. Indeed, if
the chemical composition affects Cepheid pulsation properties, this systematic
effect will reflect on the associated extragalactic distance scale.

On the other hand, the chemical composition of Cepheids can provide
information on the chemodynamical evolution of galaxies. Abundances and
abundance ratios of many chemical elements can be measured from Cepheid
spectra to derive their distribution in different environments within the Galaxy
and beyond.
For instance, abundance gradients in the Galactic disk for
25 chemical elements (from carbon to gadolinium) were measured
\citep{2002A&A...381...32A, 2002A&A...392..491A, 2004A&A...413..159A,
  2003A&A...401..939L,2007A&A...467..283L}.

The standard method for determining Cepheid atmospheric parameters and
abundances is based on one-dimensional plane-parallel hydrostatic stellar
model atmospheres. The models cover a wide range of the stellar parameters
\citep[e.g.,][]{1992IAUS..149..225K} allowing one to estimate the effective
temperature, surface gravity, and microturbulent velocity, which change with
pulsational phase, as well as the phase-independent abundances 
\citep[see e.g.][]{ 2004AJ....128..343L,
 2005AJ....129..433K, 2005AJ....130.1880A, 2008AJ....136...98L}. Physically,
the pulsational period is of the same order as the dynamical timescale 
 defined by the sound travel time across the star. The
relaxation time for mechanical disturbances caused by pulsational waves in
comparison with the pulsational period characterizes the deviation from
hydrostatic conditions. For a Cepheid model with a ten-day period, deviations
from hydrostatic conditions are expected to be on the level of a few percent
\citep{1987VA.....30..197G}; this is also the case for thermal deviations. From this
perspective one would expect that the standard approach should work rather
satisfactorily.

Dynamical non-linear stellar models of pulsating stars restricted to one
spatial dimension (1D) have a long history \citep[see
  e.g.][]{1962ApJ...136..887C, 1964RvMP...36..555C, 1966ApJ...144.1024C,
  1982ApJ...262..330S, 1994ApJS...93..233B, 1999ApJS..122..167B, 2005ApJ...632..590M, 
2008AcA....58..193S}. In those works, the convective flux
  was modelled using a 1D time-dependent theory of convection and a
  characteristic scale length over which convection mixes mass, typically
  taken proportionally to the pressure scale height.  Models were aimed to
  reproduce the light and radial velocity curves and location of the
  instability strip in the Hertzsprung-Russel diagram. Moreover,
  \cite{1991MNRAS.250..258F}, \cite{1996A&A...307..503F}, and
  \cite{1997A&A...325.1013F} used 1D models to perform spectral synthesis
  calculations to investigate shock propagation and turbulence in Cepheid
  atmospheres.

Recently, two-dimensional (2D) hydrodynamic simulations of Cepheid envelopes
were performed by \cite{2013MNRAS.435.3191M} and \cite{2015MNRAS.449.2539M},
including comparisons to convection prescriptions commonly applied in 1D
modelling. Their simulations show that the strength of the convection zone
varies significantly over the pulsation period and exhibits a phase shift
relative to the variations in radius. To successfully match this
multi-dimensional result by a 1D static, or even time-dependent convection
model, appears challenging. The authors conclude that significant improvements
are needed to make predictions based on 1D models more robust and to improve
the reliability of conclusions on the convection-pulsation coupling drawn from
them. Multi-dimensional simulations can provide guidelines for developing
descriptions of convection then applied in traditional 1D
modelling. \cite{2017A&A...600A.137F} used the CO5BOLD
radiation-hydrodynamics code to produce an exploratory grid of global three-dimensional (3D)
"star-in-a-box" models of the outer convective envelope and the inner
atmosphere of asymptotic giant branch (AGB) stars to study convection, pulsations, and shock waves and
their dependence on stellar and numerical parameters. For these stars with
low-effective temperature and gravity, the lifetime of convective cells is
comparable to the pulsational timescale, which makes it possible to model
pulsations, but the spatial resolution is not high enough for detailed
spectral synthesis. In summary, calculation of a grid of multi-dimensional
Cepheid atmosphere models is a sizable computational problem due to the
presence of different spatial and temporal scales.

Here we present the -- to our knowledge -- first spectroscopic investigation
of a 2D envelope models  of a Cepheid-like variable where the non-local and
time-dependent nature of convection is included from first principles.
Radiative transfer is treated non-locally, albeit in grey
approximation. Our intention is to demonstrate that the model is able to
capture relevant features of the Cepheid physics despite the employed
approximations. In Section~2 we describe the model, in Sect.~3 we study its
spectroscopic properties. Specifically, we derive the microturbulent velocity
from the 2D model and report on the methodology behind its derivation, as
well as describe the behaviour of the line equivalent widths and
asymmetries. In Sect.~4 we discuss line shifts and give an interpretation of
the residual spectroscopic line-of-sight radial velocity.  A derivation of the
projection factor (p-factor), which is a key parameter in the Baade-Wesselink
method of the distance determination to Cepheids, is discussed in
Sect.~5. Section~6 lists our conclusions.

\section{Description of the model}

Radiation-hydrodynamics simulations of a short-periodic Cepheid-like variable
employing grey radiative transfer were calculated with the CO5BOLD code
\citep{2012JCoPh.231..919F} in 2D Cartesian geometry. The models gt56g20n01 to gt56g20n04 have
a nominal effective temperature of $T_\mathrm{eff}=5600$\,K, and a constant
depth-independent gravitational acceleration of $\log g =2$. The
pulsation period of the Cepheid model comes out to $\approx$2.8\,days. The simulations use an equidistant spatial grid of
$\mathrm{N_x} \times \mathrm{N_z} = 600 \times$ 500 (model gt56g20n01)
to 580 (gt56g20n04)
points along the horizontal and vertical direction,
respectively.  The corresponding geometrical size of the largest model is $\mathrm{l_x} \times
\mathrm{l_z} = 1.12 \cdot 10^{12}$\,cm $\times$ $5.4 \cdot 10^{11}$\,cm. The
 size of the grid cells  is $1.88 \cdot 10^9$\,cm $\times$ $9.38\cdot 10^8$\,cm. For the
flow, impenetrable boundaries are imposed at the top and bottom of the domain; the
side boundaries are periodic. A radiative flux according to the nominal
effective temperature is prescribed at the bottom. Radiation can 
pass unimpeded through the top of the modelled box.
The chemical composition of the model is close to solar. Specifically, the
hydrogen abundance is $\log\epsilon_\mathrm{H}=12.0$, helium abundance
$\log\epsilon_\mathrm{He}=11.0$, silicon abundance
$\log\epsilon_\mathrm{Si}=7.5$, and iron abundance
$\log\epsilon_\mathrm{Fe}=7.46$.

\begin{figure}
\includegraphics[width=8.8cm]{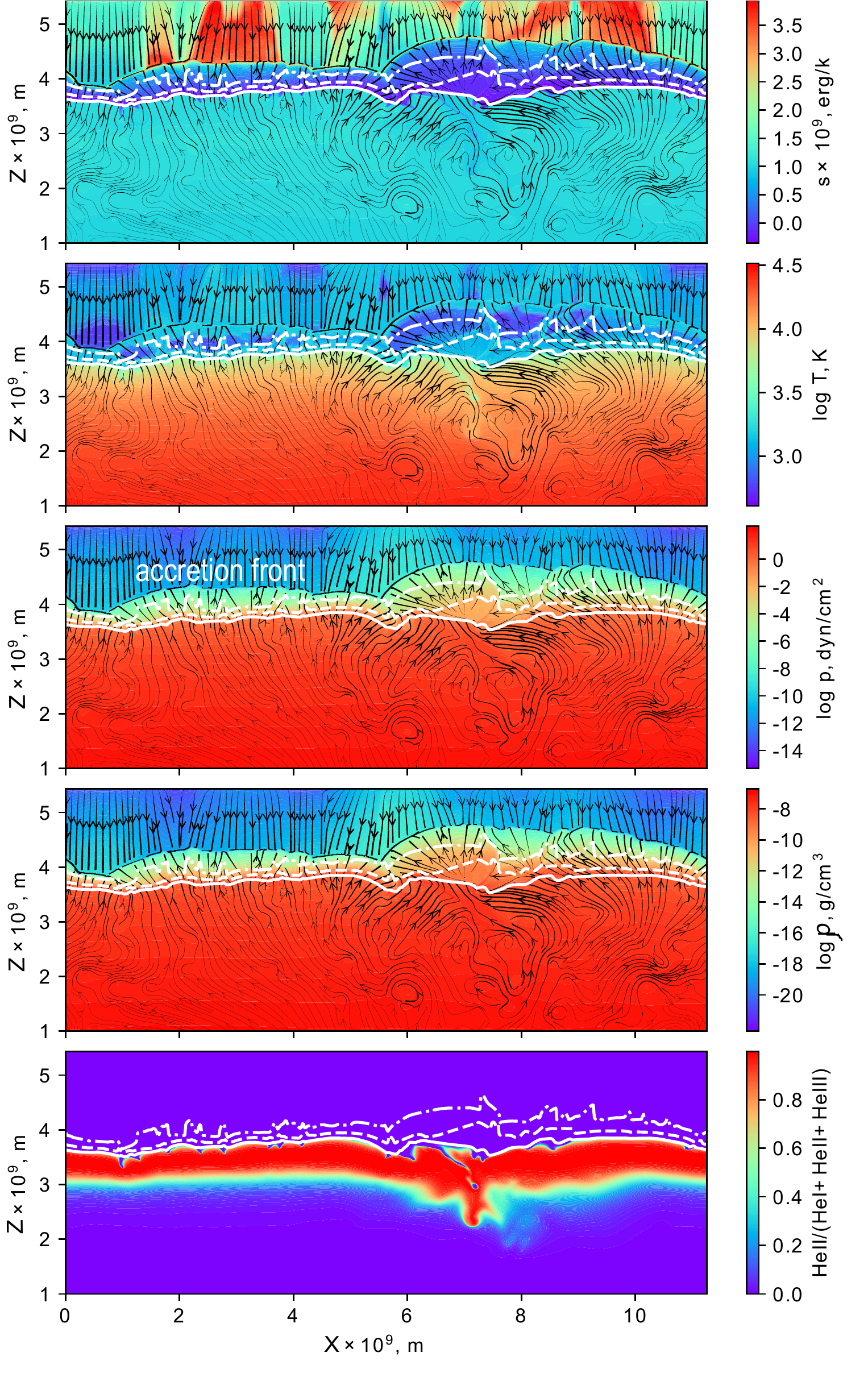}
\caption{Entropy (top panel), temperature, pressure, density, and \ion{He}{ii}
  ionization fraction
  $n_\mathrm{\ion{He}{ii}}/(n_\mathrm{\ion{He}{i}}+n_\mathrm{\ion{He}{ii}} +
  n_\mathrm{\ion{He}{iii}})$ (bottom panel) as function of the horizontal and
  vertical coordinates at time $t=1.40 \times 10^7$\,s, which corresponds to an
  expanding phase. Pseudo-streamlines are shown as black solid lines.
  Surfaces of constant Rosseland optical depth are shown by white lines:
  $\tau_\mathrm{R}=1$ (solid line), $\tau_\mathrm{R}=10^{-2}$ (dashed
  line), $\tau_\mathrm{R}=10^{-4}$ (dashed-dotted line). }
\label{2dentrop}%
\end{figure}

 \begin{figure}
\includegraphics[width=\hsize]{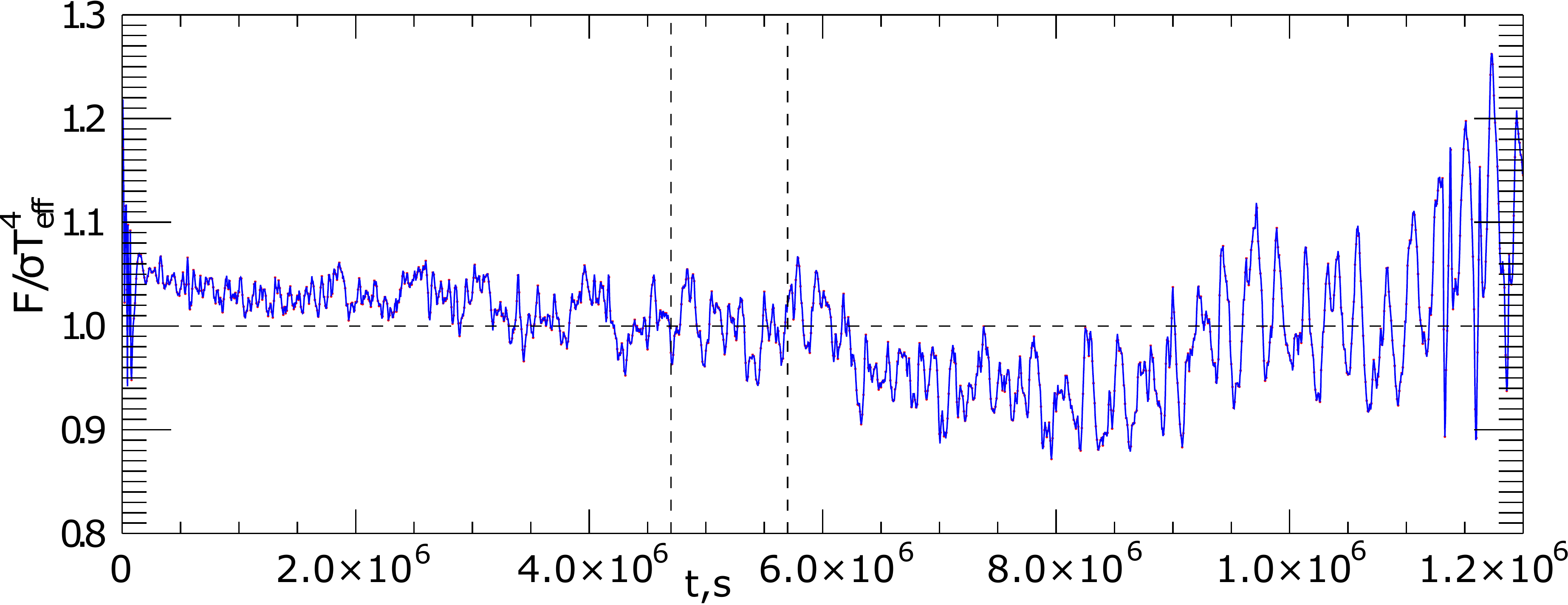}
\caption{Initial evolution in time of the bolometric flux of the model.
  Pulsations set in at $t\approx 10^6$\,s. Dotted lines indicate the time
  interval without pulsations for which spectral synthesis calculations were
  performed.}
\label{frad_initial_evolution}%
\end{figure}

\begin{figure}
\includegraphics[width=8.5 cm]{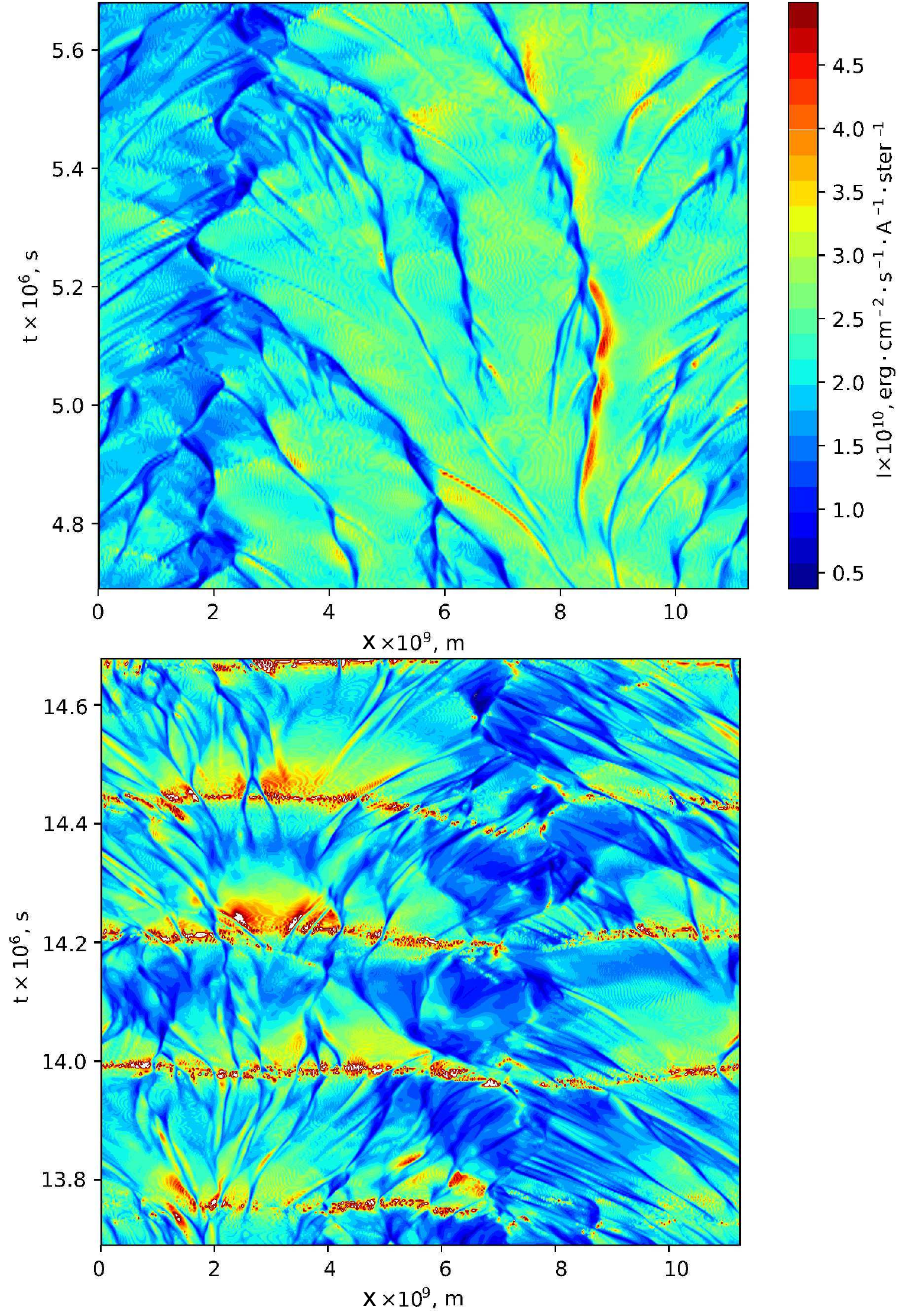}
\caption{Spatio-temporal evolution of the emergent intensity directed in
  vertical direction ($\mu = 1$). The upper panel shows a time interval when
  pulsations have not set in (model gt56g20n01), the bottom panel a time interval with fully
  developed pulsations (model gt56g20n04). Time, spatial, and colour scales are identical for both
  panels.}
\label{intmap}%
\end{figure} 
   
Figure~\ref{2dentrop} depicts images of entropy, temperature, pressure,
density, and \ion{He}{ii} ionization fraction at an instance in time during the
expansion phase as a function of geometrical coordinates. Convection produces
horizontal inhomogeneities so that surfaces of constant optical depth are not
flat. The pulsating model exhibits strong shocks. Low density gas is falling
down from above and collides with expanding photospheric and sub-photospheric
material. At the so-formed accretion front,  density and pressure jump by up to
three orders of magnitude.  We checked that such jumps cannot be simply
understood as the result of the Rankine–Hugoniot (adiabatic) shock conditions but
further factors must play a role. As evident from the lines of constant
optical depth, line formation typically takes place in the region
between the accretion front and optical depth unity.  

We emphasize that the construction of the model was a numerical
challenge.
The largest obstacle is the extremely small numerical time step
imposed by the short radiative relaxation time on a time-explicit scheme
\citep[for a detailed discussion, see][]{2013MNRAS.435.3191M}.
It enforces the restriction to 2D models, at the moment.
In addition, strong pulsations can tend to cause an imbalance 
between mass in- and outflow resulting in a net time-averaged mass flow 
across the top boundary. To avoid it, the closed top boundary was chosen.
However, the closed top boundary enhances strong artificial 
velocity gradients close to the boundary. To decrease the influence 
of them on dynamics of optically thin regions, we applied an artificial
drag force in a number of grid layers close to the top, reducing the
velocities by a certain fraction per time interval. This numerical
approach impacts on the dynamics of the line formation regions of strong lines,
which  like the Ca triplet form at the outer atmosphere near the chromosphere. 
To avoid the impact of boundary effects we consider lines that are forming deeper in  the range $\log \tau_\mathrm{R}=-3 \ldots 0$.

The model shows self-excited fundamental mode pulsations starting from
initially plane-parallel static conditions (adding random small amplitude
disturbances to break the symmetry). The initial evolution in time  of the bolometric flux of the model is shown in
Fig.~\ref{frad_initial_evolution}. Pulsations set in  at around $\approx 6 \times
10^6$\,s. They are  excited by the
$\kappa$-mechanism \citep{1917Obs....40..290E, 1963ARA&A...1..367Z}.  In
particular, around 4\,\%  of the radiative flux $\sigma T^4_\mathrm{eff}$ is
spent to ionize hydrogen, which is fully ionized for Rosseland optical depth
$\tau_\mathrm{R}>1$, due to the steep temperature gradient, and is largely neutral
for $\tau_\mathrm{R}<1$.  The main driver of pulsations is the region of
singly ionized helium (see bottom panel in Fig.~\ref{2dentrop}). The thickness
of the \ion{He}{ii} ionization zone varies; \ion{He}{ii} ionization stores and
releases~16\,\% of the radiative energy passing the layer during a
pulsational cycle.

Before pulsations have set in, the flow evolution is mostly governed by
convection.  Regions with up- and down-flows at the photosphere affect the
outgoing radiation field.  The spatio-temporal evolution of the emergent
intensity (in vertical direction, $\mu=1$) when pulsations have not set in is
shown in the top panel of Fig.~\ref{intmap}. The intensity map exhibits a
typical convective pattern. In particular, convection generates a few large-scale down-flows, which exist for long timescales $> 10^6$\,s. Smaller-scale
down-flows of cooled gas have a tendency to get ``sucked'' into the dominant
larger-scale flows. Down-flows correlate with lower intensities, the bluer regions in Fig.~\ref{intmap}. 
 
The intensity map for later phases when pulsations have set in is shown in the
bottom panel of Fig.~\ref{intmap}.  The convective pattern now consists of a
dominant large-scale down-flow in the right part of the intensity map, and
small-scale flows that evolve independently.  For the given
atmospheric parameters and assuming a mass of roughly 5 \(\textup{M}_\odot\)  , one can expect around 300 granulation cells on the visible hemisphere. The mass was motivated from inspection of 
the PAdova and TRieste Stellar Evolution Code (PARSEC) evolutionary tracks  \citep{2012MNRAS.427..127B}.

\begin{figure}
\includegraphics[width=9cm, scale=0.7]{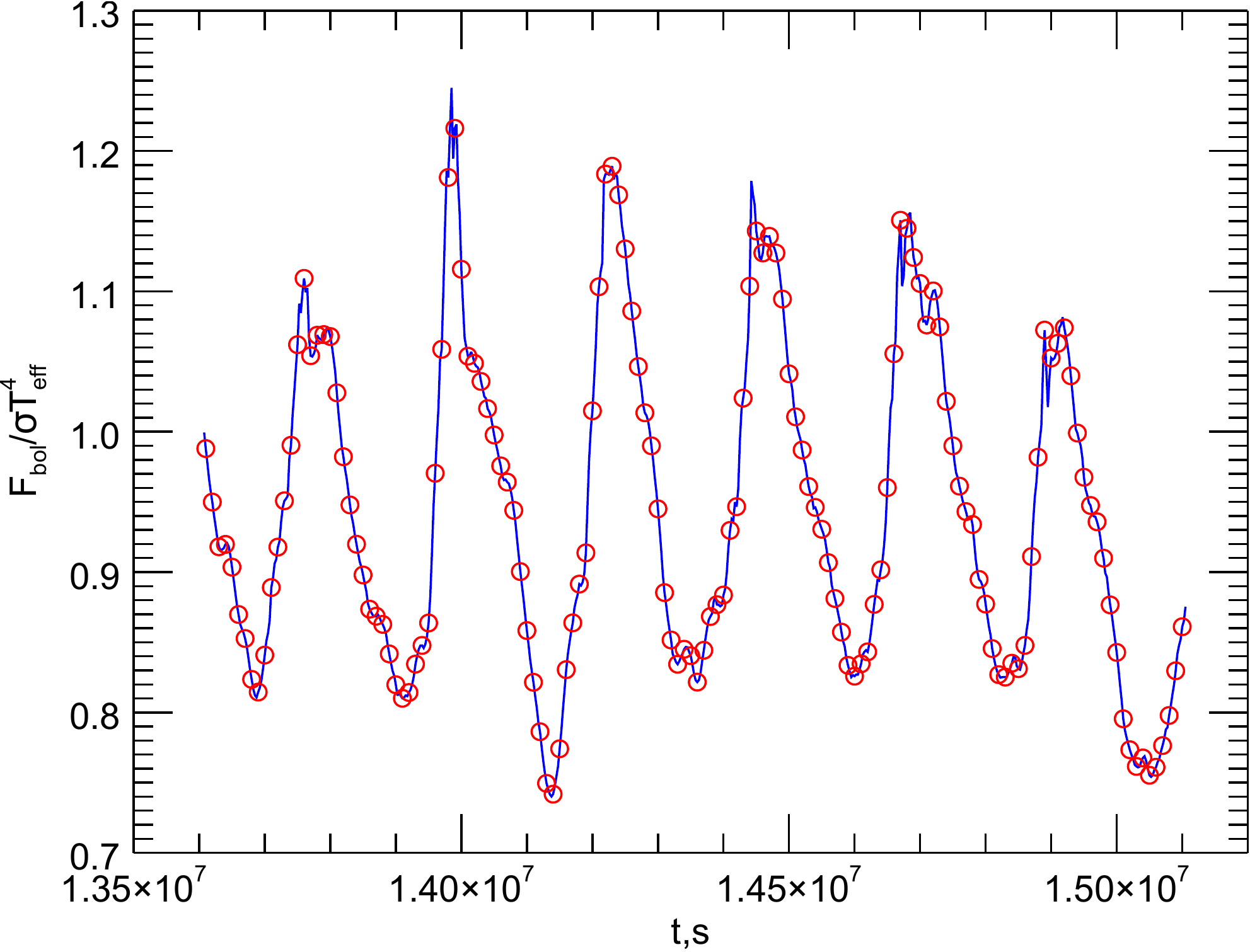}
\caption{Model light curve in terms of the emergent bolometric flux. Red
  circles mark the instances in time for which a spectral synthesis was
  performed.}
\label{LC}%
\end{figure}

The phase of maximum compression
roughly coincides with the phase of maximum brightness (see also
Fig.~\ref{LC}). Maximum brightness is not reached at exactly the same time at
all locations but depends slightly on the morphology of the convective surface
flow. For instance, in the large down-flow maximum brightness is reached
slightly earlier than in the neighbouring regions. This illustrates the impact
of horizontal inhomogeneities on the pulsations.

\begin{figure}
\includegraphics[width=8.5 cm]{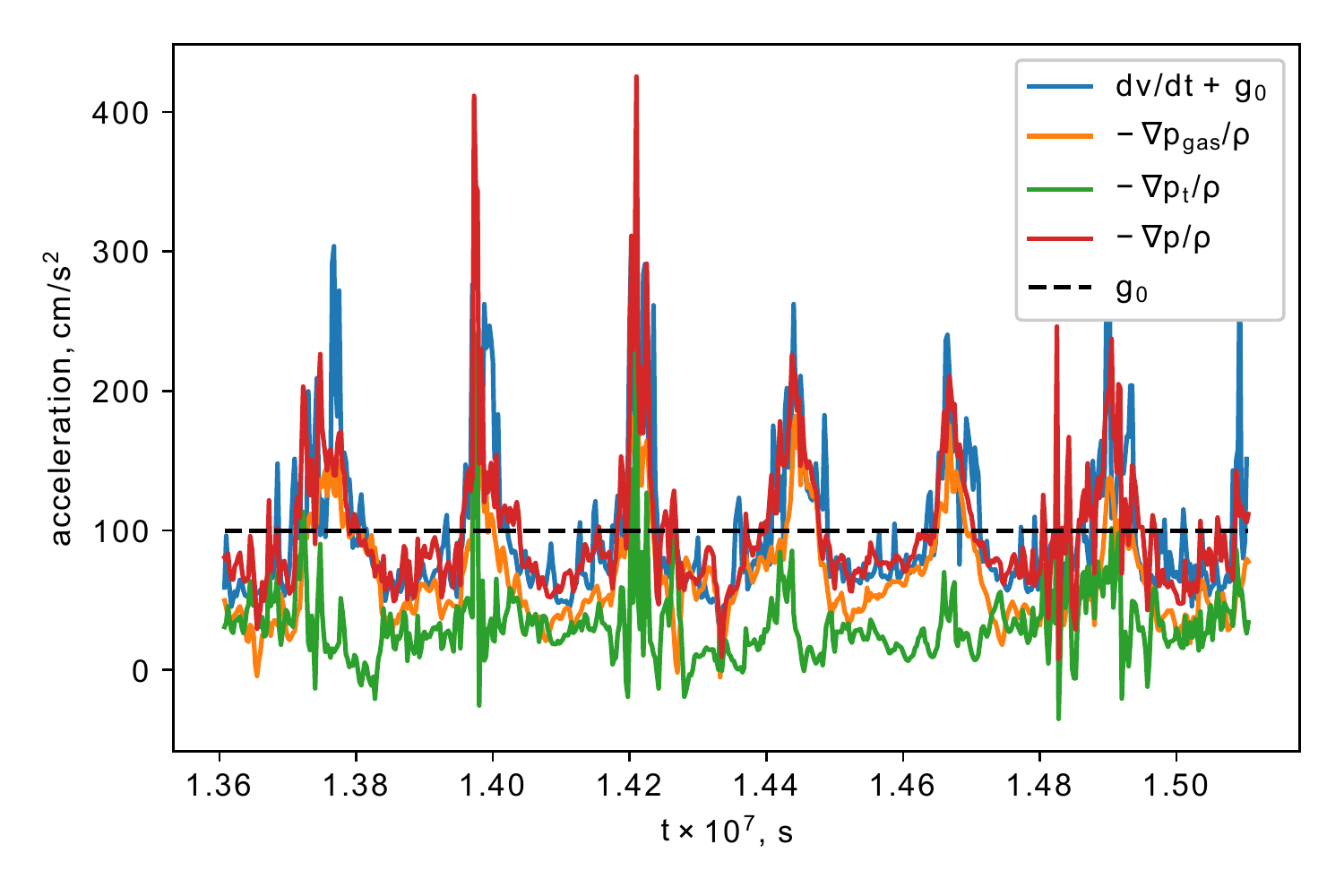}

\caption{The total acceleration $dv/dt + g_0$ experienced by a
  (Lagrangian) mass element close to optical depth unity as a function of
  time,  where  $g_0$ is the constant
  gravitational acceleration. Coloured lines show the contributions to the
  pressure gradient by the gas
  pressure~$p_\mathrm{gas}$, and turbulent pressure~$p_\mathrm{t}$, as well
  as the resulting sum~$p=p_\mathrm{gas} + p_\mathrm{t}$ balanced by the
  total acceleration.}
\label{accel}%
\end{figure}

Preferentially during phases of maximum brightness, which almost 
coincide with the  maximum compression, convective down-flows are generated. This is a sign of the action
of the convective instability.  The efficiency of convection depends on the
gravitational acceleration $g_0$, and the growth timescale of the convective
instability~$\tau_\mathrm{conv}$ is proportional $\tau_\mathrm{conv} \sim
1/\omega_\mathrm{BV} \sim \sqrt{1/g_0}$ , where $\omega_\mathrm{BV}$ is the
Brunt–V\"ais\"al\"a frequency.  The time evolution of the acceleration $a$ of
a pseudo-Lagrangian mass element following the mean vertical mass motion close
to the photosphere is shown in Fig.~\ref{accel}.  The model itself is based on
an Eulerian description of the flow field from which one can transform to a
pseudo-Lagrangian reference frame.  During the reversal of the direction of
motion at the phase of maximum compression, a photospheric pseudo-Lagrangian
mass element experiences an acceleration of $\approx (1.0 \ldots 2.5) g_0$.  In
total it experiences an effective acceleration of $g_\mathrm{eff}=a+g_0
\approx (2.0 \ldots 3.5) g_0$ , which amplifies the convective instability and
also leads to high convective velocities. One can see in Fig.~\ref{accel} that
the turbulent pressure $p_\mathrm{t}$ adds a significant contribution to the
total pressure gradient, which is equal to the effective gravity: $-\nabla (
p_\mathrm{gas} + p_\mathrm{t})/\rho=a+g_0$.

\begin{figure}
\includegraphics[width=9cm, scale=0.7]{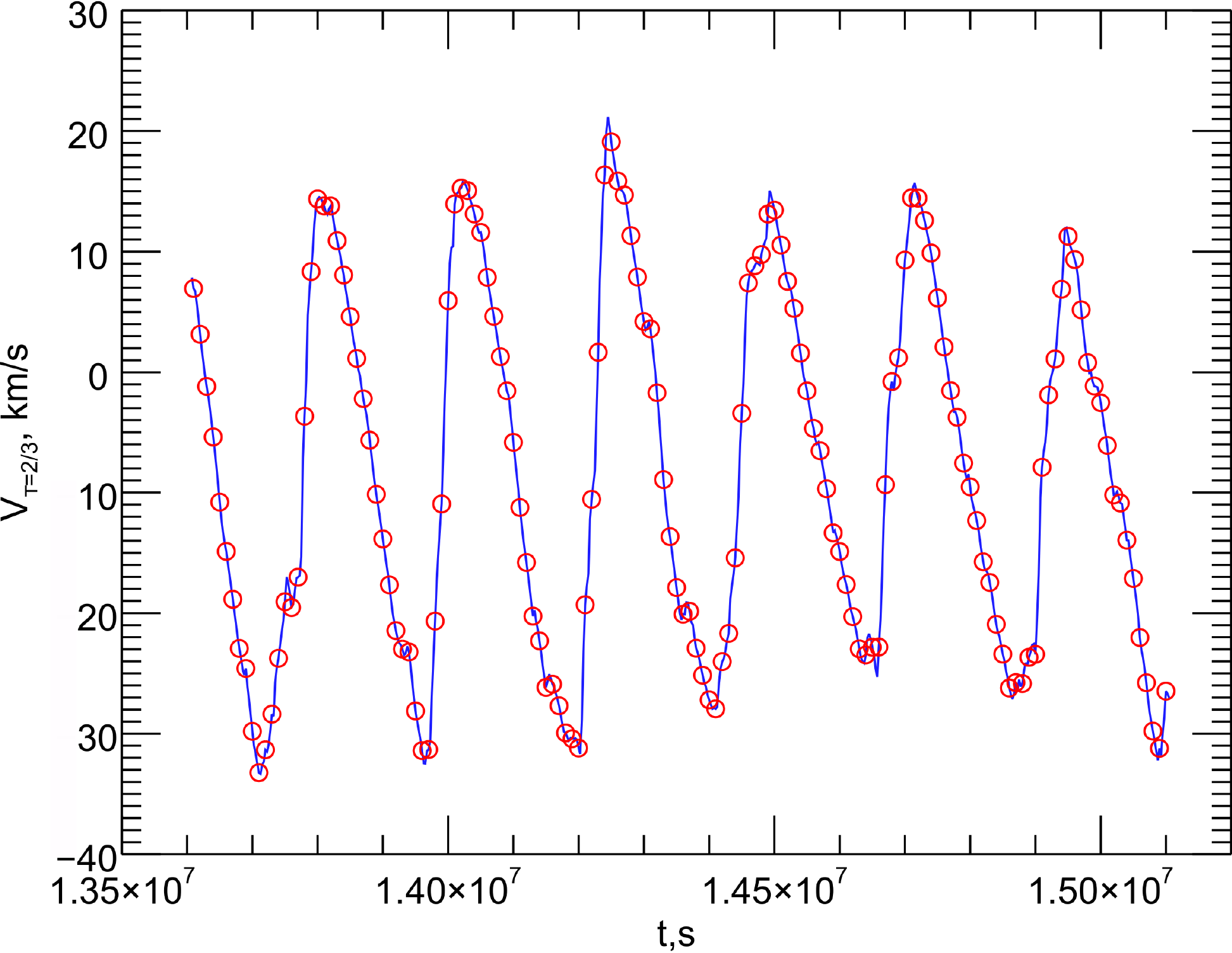}
\caption{Vertical (radial) velocity of a layer at fixed Rosseland optical
  depth $\tau_R=2/3$. Here, a positive velocity corresponds to a blueshift.}
\label{RVC}%
\end{figure}

A 1D mean model, which is the result of horizontal averaging of the full 2D
model at fixed geometrical height, was used to analyze the pulsational
velocity and radiation flux.  The radial velocity curve of a layer with the
Rosseland optical depth $\tau_\mathrm{R}=2/3$ shown in Fig.~\ref{RVC} was
calculated by cubic spline interpolation of the vertical velocity component of
the mean 1D model. The amplitude of the radial velocity is $V \approx 20$\kmos
, which is quite typical for a Cepheid \citep{2013A&A...550A..70G}. The  non-zero mean vertical velocity 
of the optical surface is the consequence of the 
presence of convective fluctuations, 
large amplitudes of pulsations and a non-linearity of opacities as a function 
of the temperature.
A word on sign conventions: in the
CO$^5$BOLD simulations the vertical axis is directed toward the observer so
that a positive vertical velocity corresponds to a blueshift.  However,
normally the spectroscopic convention for the radial velocity is used in this
work, where a blueshift corresponds to a negative velocity. 
 
A light curve in terms of the bolometric flux is shown in Fig.~\ref{LC}. Red
circles mark instances in time for which a spectral synthesis was performed
with Linfor3D\footnote{http://www.aip.de/Members/msteffen/linfor3d/.} \citep{2016arXiv161004427G}.  The
light curve covers a time interval with fully developed pulsations covering
six periods. For each 2D snapshot, Linfor3D solves the radiative transfer
equation along three inclined angles for two azimuthal angles, which are lying
in the xz-plane, and along the vertical direction.  To accelerate the spectral
synthesis one can sub-sample the model structure along the x-axis.
Tests taking each and every third grid point along the x-axis were
performed for a line \ion{Fe}{i} $\lambda6003$\,\AA. 
Equivalent widths for
these cases come out very similar: $W_\mathrm{\Delta n_{x}=1}=109.8$\,m\AA\ 
and $W_{\mathrm{\Delta n_{x}}=3}=111.2$\,m\AA. Due to the similarity, the
computationally less demanding sub-sampled case is used in this paper.

\subsection{Convective noise}

Observed light and radial velocity curves of Cepheids are typically smooth,
and show a high degree of similarity from cycle to cycle.  The model light
curve in Fig.~\ref{LC} and the radial velocity curve in Fig.~\ref{RVC} do not
at all exhibit these properties.  Amplitude and shape of the curves change
significantly during the pulsations. This is the result of convection, which
adds statistical fluctuations to the velocity and thermal structure of the
model. The fluctuations in the pressure gradients shown in Fig.~\ref{accel}
are a clear illustration of this ``convective noise''.  The convective noise
appears so prominent in our light and velocity curves since our model
represents only a small part of the surface of a Cepheid. The fluctuations
would decrease if we increased the spatial extent of the model, or could take
recourse to many realizations of the stellar surface with independent
convective but coherent pulsational properties. However, both options would
demand prohibitively expensive additional simulations. In the following we try
to mitigate the effects of the fairly low resulting signal-to-noise ratio by
adequate procedures when extracting simulation properties.  As a side, we note
that the cycle-to-cycle variations in Cepheids recently reported by
\citet{2016MNRAS.463.1707A} might be an imprint of the residual convective
noise after averaging over the full stellar disk. However,  it is difficult
to extrapolate the convective noise properties of our calculated model to  
a long-period Cepheid. According to the period-mean gravity relation 
\citep{1965ApJ...142.1649G}, it requires approximately a  factor of ten lower
gravitational  acceleration  for a 35-day pulsation period, 
which changes the convective  timescale, velocities,  
characteristic sizes, and lifetimes of convective cells.

\subsection{Effects of the Cartesian geometry and grey radiative transfer}

Our Cepheid model has Cartesian geometry while the structure of a real Cepheid
is more accurately represented by a spherical geometry. We want to estimate the
impact of the Cartesian geometry here. 

In Cepheids the radius changes due to the pulsations are of the order of
$\approx 10 \%$, which corresponds to a change in the stellar volume of
$\left( \frac{\Delta V}{V}\right)_\mathrm{s}=3\frac{\Delta R}{R}=0.3$, where
the subscript~s indicates a spherical geometry.  For the planar (subscript~p)
geometry adopted here, the change is $\left( \frac{\Delta V}{V}
\right)_\mathrm{p}=\frac{\Delta R}{R}=0.1$.  For adiabatic pulsations, there
is a corresponding difference of the temperature change $\Delta \frac{\Delta
  T}{T} = \left[ \left( \frac{\Delta V}{V} \right)_\mathrm{s} - \left(
  \frac{\Delta V}{V} \right)_\mathrm{p} \right] \cdot
(\gamma_\mathrm{ad}-1)=0.2 \cdot (\gamma_\mathrm{ad}-1)$, where
$\gamma_\mathrm{ad}$ is the adiabatic exponent, which is equal to $5/3$ for an
ideal monatomic gas. One would arrive at the same conclusion if one considered
an atmosphere with a fixed thickness during the pulsations.  Hence, the
different geometry leads to a temperature difference of $\approx 13 \%$. This
is significant, and eventually means noticeable changes in the conditions for
line formation. However, as we will see in a moment, the temperature does not
follow adiabatic changes but is rather controlled by radiation. Radiative
properties differ less between the geometries since the thickness of the
atmosphere is still rather small in a Cepheid -- significantly less than its
radius change during the pulsational cycle.
  
\begin{figure}
\includegraphics[width=8.5 cm]{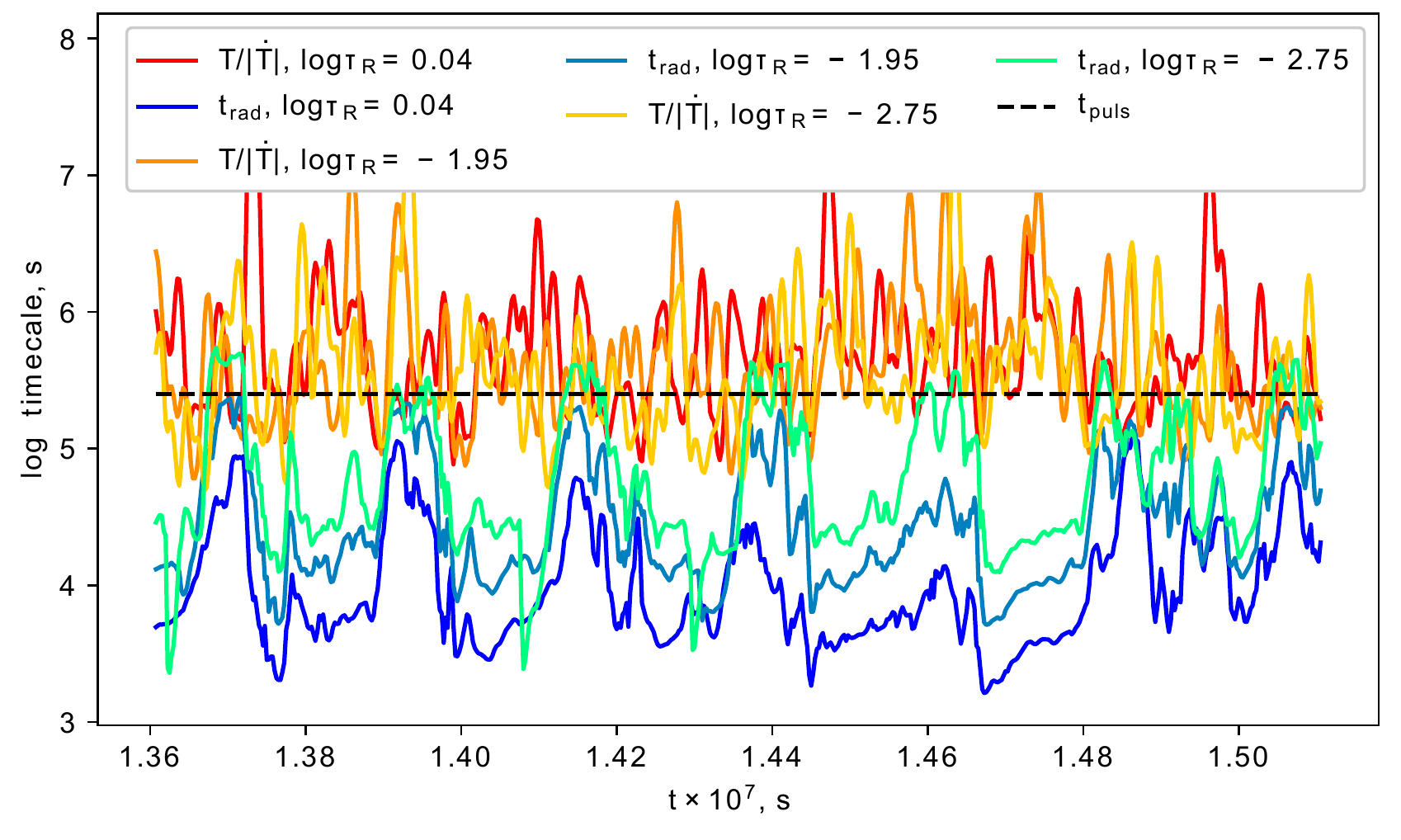}
\caption{Comparison of the characteristic radiative relaxation time and the
  actual timescales on which temperature changes happen for three Lagrangian
  mass elements in the atmosphere. The elements are located in the line
  formation region at time-averaged optical depths
  $\log\tau_\mathrm{R}=0.04,-1.95,-2.75$.}
\label{timescales}%
\end{figure} 
   
One can argue that using grey radiative transfer in the model is far from
realistic and may lead to different qualitative changes of the thermal
structure during the pulsations in comparison to the case of
frequency-dependent (non-grey) radiative transfer. Especially the
characteristic timescale of radiative relaxation, which is defined as the
time for reaching radiative equilibrium conditions in the atmosphere, depends
on the opacity treatment. To derive characteristic timescales we follow
\cite{2002A&A...395...99L}. With the assumption that the geometrical size of a
thermal disturbance is equal to the local pressure scale height, the radiative
relaxation timescale in Eddington approximation can be written as:
\begin{equation}
    t_\mathrm{rad}=\frac{c_\mathrm{p}}{16 \sigma \chi T^3} \Big( 1+\frac{3}{4 \pi^2} \tau^2_\mathrm{dis}\Big ),
\end{equation}
where $c_\mathrm{p}$ denotes the specific heat at constant pressure, $\sigma$
Stefan-Boltzmann's constant, $\chi$ the Rosseland opacity, and
$\tau_\mathrm{dis}=\chi \rho H_\mathrm{p}$ the optical size of the assumed
temperature disturbance.

The actual characteristic time of change of the temperature $t_\mathrm{T}$ of
a mass element in the atmosphere model can be calculated as 
\begin{equation}
    t_\mathrm{T}=\frac{T}{|dT/dt|}, 
\end{equation}
where $dT/dt$ is a rate of the temperature change.  This timescale depends on
local fluctuations of the temperature due to the presence of disturbances
generated by convection or pulsations.  The pulsational period
$t_\mathrm{puls}$ characterizes the global change of the thermal structure
during the pulsations.

Figure~\ref{timescales} shows the timescales as a function of time for three
Lagrangian mass elements in the atmosphere, which are lying in the line
formation region with time-averaged optical depths
$\log\tau_\mathrm{R}=0.04,-1.95,-2.75$.  The properties of the mass elements
were calculated from the 1D mean model. The radiative relaxation timescale is
shorter than both the pulsational period and the characteristic time of the
temperature change in the line formation regions, except in phases of maximum
compression, during which for the highest parts of formation regions the timescale is comparable with the period of pulsations.  One can see that the
characteristic time of change of the temperature does not depend much on the
optical depth for the line formation region and it has the same order as the
pulsational period.  The radiative timescales have been evaluated under
simplifying assumptions, and should therefore be taken as order of magnitude
estimates only.  Our choice of the disturbance size is rough. One might argue
that the size of horizontal disturbances generated by convection is a more
appropriate length scale, and typically amounts to a few pressure scale
heights. Even with $10H_\mathrm{P}$ $, t_\mathrm{rad}$ does not change much
since the optical depth is already quite low (in the optically thin limit the
radiative timescale is independent of the size of the disturbance). We
conclude that radiation is able to keep the temperature structure close to
radiative equilibrium -- consistent with actual modelling results for the line
formation region. For non-grey radiative transfer, the radiative timescale is
shorter \citep[see e.g.][]{2002A&A...395...99L}. The stronger coupling
between radiation and matter in the non-grey case would maintain radiative
equilibrium conditions even more effectively. However, since the grey
radiative transfer is already capable of doing so, the main difference to the
non-grey case would be the lack of line blanketing and back warming, which is
rather modest. One also expects  in the non-gray case  
smaller temperature fluctuations in the photosphere,
and   the mean temperature gradient might be steeper.
We conclude that our 2D Cartesian model provides fairly
realistic estimates of the thermal conditions in the line formation region.

\subsection{Relating velocity and luminosity amplitudes}
 \cite{1995A&A...293...87K} derived a universal relation
 between the luminosity $(\delta L/L)_\mathrm{bol}$ and velocity $v_\mathrm{osc}$
 amplitudes  of  oscillations for many classes of oscillating stars using linear
 theory and observational data:
\begin{equation}
(\delta L/L)_\mathrm{bol} \sim \frac{v_\mathrm{osc}}{ T_\mathrm{eff}}.
\end{equation}
The relation was calibrated using $\beta$ and $\delta$ Cephei, $\delta$ Scuti, and RR Lyrae stars.  For a
given $T_\mathrm{eff}$  and $v_\mathrm{osc}$ , one can 
predict the luminosity amplitude $(\delta 
L/L)_\mathrm{pred}$ and compare it with the observed 
luminosity amplitude. For the 2D Cepheid model, the predicted 
luminosity amplitude $(\delta L/L)_\mathrm{pred} \approx 0.375$ for the  $v_\mathrm{osc} \approx 20$\kmos and $T_\mathrm{eff}=5600$ K. The "observed"  
luminosity amplitude of the model is $\approx0.35$. Both of them are in good 
agreement with the result  of  \cite{1995A&A...293...87K} within the  $1\sigma$ scatter.

\section{Spectroscopic properties}
Spectral properties are directly accessible to observation, and we want to
characterize our model in terms of its spectroscopic characteristics.
Microturbulence is an important "side parameter" in a standard spectroscopic
analysis.  \cite{1999A&A...344..935G} derived the microturbulence velocity
curve of $\delta$ Cep using  1D non-linear non-adiabatic pulsating
models. Perhaps not surprisingly, the microturbulent velocity turns out to
change with pulsational phase. Here we attempt a first pilot study of the
microturbulence derived from a multi-dimensional Cepheid model.

\subsection{Microturbulence}

Our estimate of the microturbulent velocity is based on the method described
in \cite{2013MSAIS..24...37S}. The procedure is as follows: For a spectral
line, a spectral synthesis is performed using the 2D thermal structure with
different velocity fields: (i) using the original 2D velocity field
$v_\mathrm{2D}$, and (ii) replacing the 2D hydrodynamic velocity field with an
isotropic, depth independent microturbulent velocity $\xi_\mathrm{mic}$ as in
a case of classical 1D spectral synthesis.  The microturbulent velocity for
the considered spectral line at each instance in time is defined by the
requirement $\mathrm{EW}(v_\mathrm{2D})=\mathrm{EW}(\xi_\mathrm{mic})$.  In
practice, we calculated $\mathrm{EW}(\xi_\mathrm{mic})$ for a small grid of
$\xi_\mathrm{mic}$ and interpolated in this grid to the desired
$\mathrm{EW}(v_\mathrm{2D})$. This method allows us to determine a
microturbulent velocity for each individual line. It is clear that it differs
from the classical procedure where one considers a set of weak and strong
lines. However, besides having the virtue of being able to assign a
microturbulence to a single line, the procedure applied here is perhaps
the cleanest possible way to derive a microturbulence if one wants to actually
associate the microturbulence with a velocity field providing the non-thermal
line broadening.

A set of fictitious neutral and singly ionized iron lines with fixed
wavelength $\lambda=5500$\,\AA\ but different line strengths and excitation
potentials was used to conduct a systematic survey of the microturbulent
velocity as a function of time. The excitation energies for \ion{Fe}{i} lines
were $E_i=1,3,5$\,eV; for \ion{Fe}{ii} they were $E_i=1,3,5,10$\,eV.  Oscillator
strengths were chosen for each excitation energy $E_i$ so as to cover a range
of equivalent widths from 10\,m\AA\ to 130\,m\AA\ in steps of 20\,m\AA\ for
the initial instance in time at step~(i).

\begin{figure}
\includegraphics[width=9.7cm, scale=0.6]{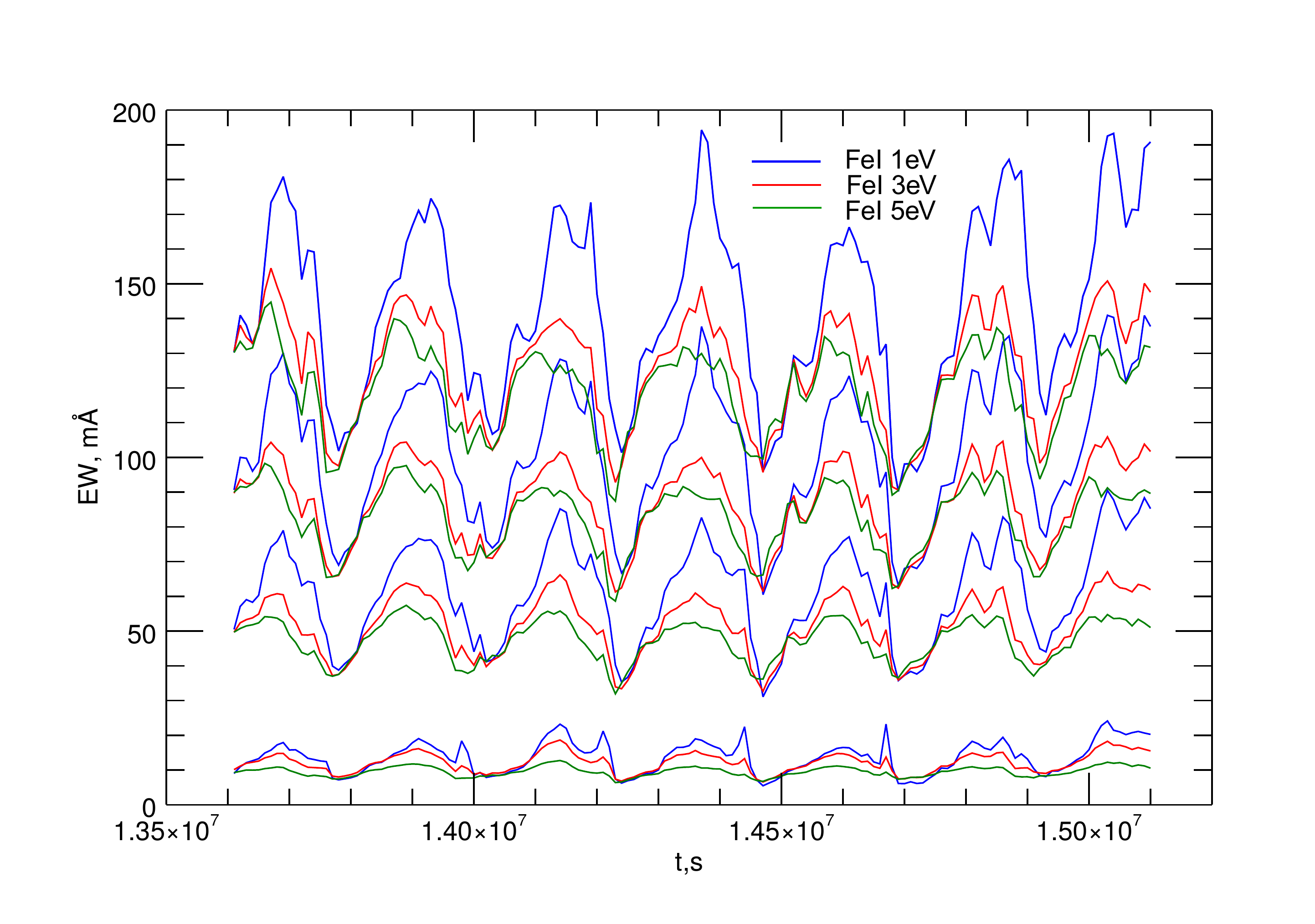}
\caption{Equivalent widths obtained from the 2D model as a function of time
  for lines of \ion{Fe}{i} $\lambda5500$\,\AA\ with excitation energies of
  $E_i=1,3,5$\,eV (coloured solid lines). Different oscillator strengths were
  used for each excitation energy to cover a range of equivalent widths from
  10\,m\AA\ to 130\,m\AA\ at starting time, as apparent by the four groups
  of three lines each.}
\label{FeIew}%
\end{figure}

\begin{figure}
\includegraphics[width=9.1cm, scale=0.6]{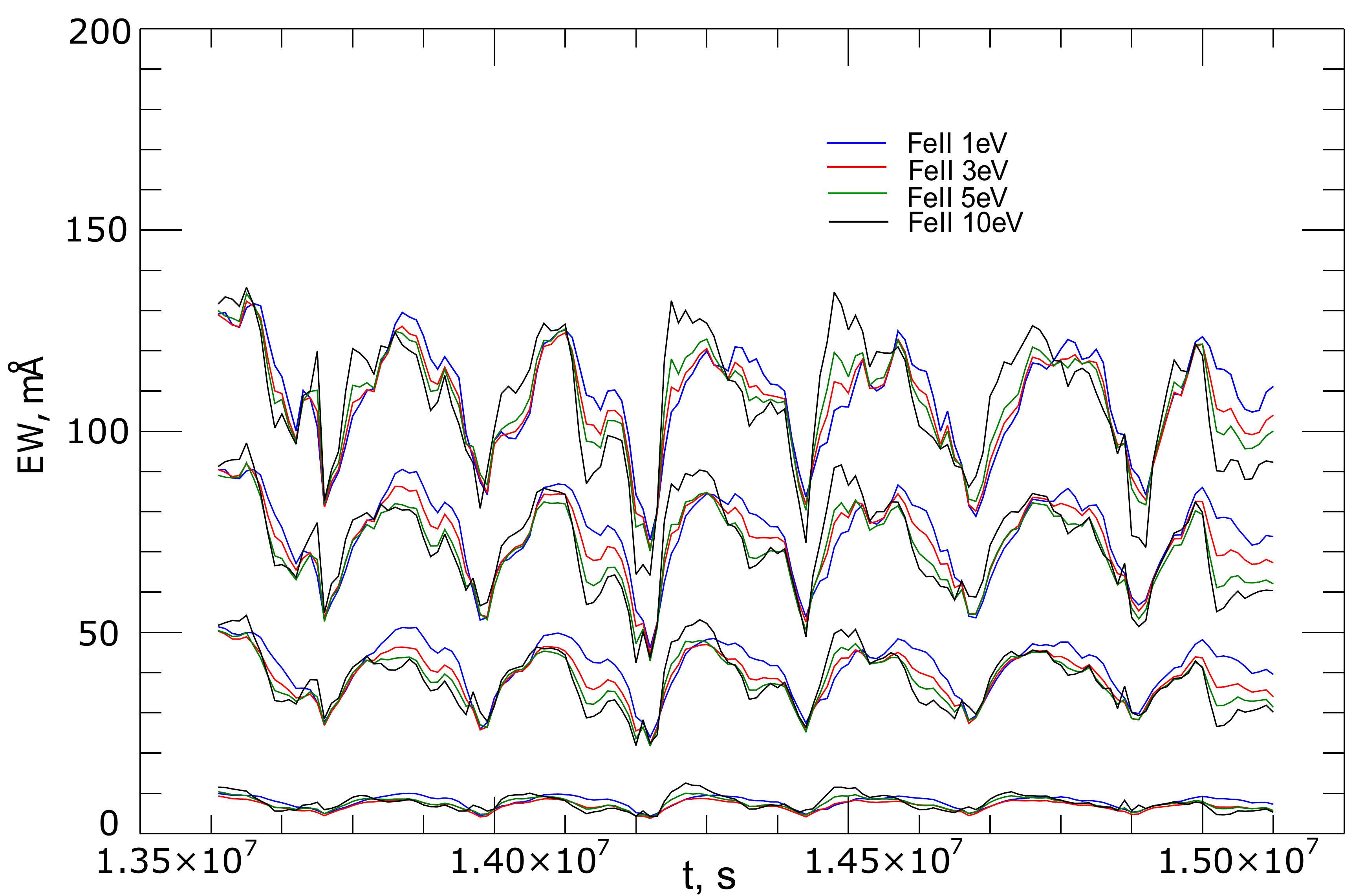}
\caption{As Fig.~\ref{FeIew} but for lines of \ion{Fe}{ii} $\lambda5500$\,\AA\
  with $E_i=1,3,5,10$\,eV.}
\label{FeIIew}%
\end{figure}

Based on the 2D model, equivalent widths (EWs) for a set of \ion{Fe}{i} and
\ion{Fe}{ii} lines as a function of time in step (i) are shown in
Figs.~\ref{FeIew} and~\ref{FeIIew}.  Both groups of these lines, \ion{Fe}{i}
and \ion{Fe}{ii}, show a modulation of the equivalent width with
pulsational phase for both strong and weak lines. The amplitude of the EW
variation is bigger for \ion{Fe}{i} than for \ion{Fe}{ii}, for the same
initial line strengths.  Lines of the \ion{Fe}{i} and \ion{Fe}{ii} with bigger
line strength have higher amplitude of the variation of EWs.  The overall
behaviour can be directly understood from the different temperature sensitivities
of the lines in the two different ionization stages. Besides a roughly
periodic variation of the equivalent widths, there is a stochastic component
stemming from the convective noise in the model.

\begin{figure}
\includegraphics[width=\hsize]{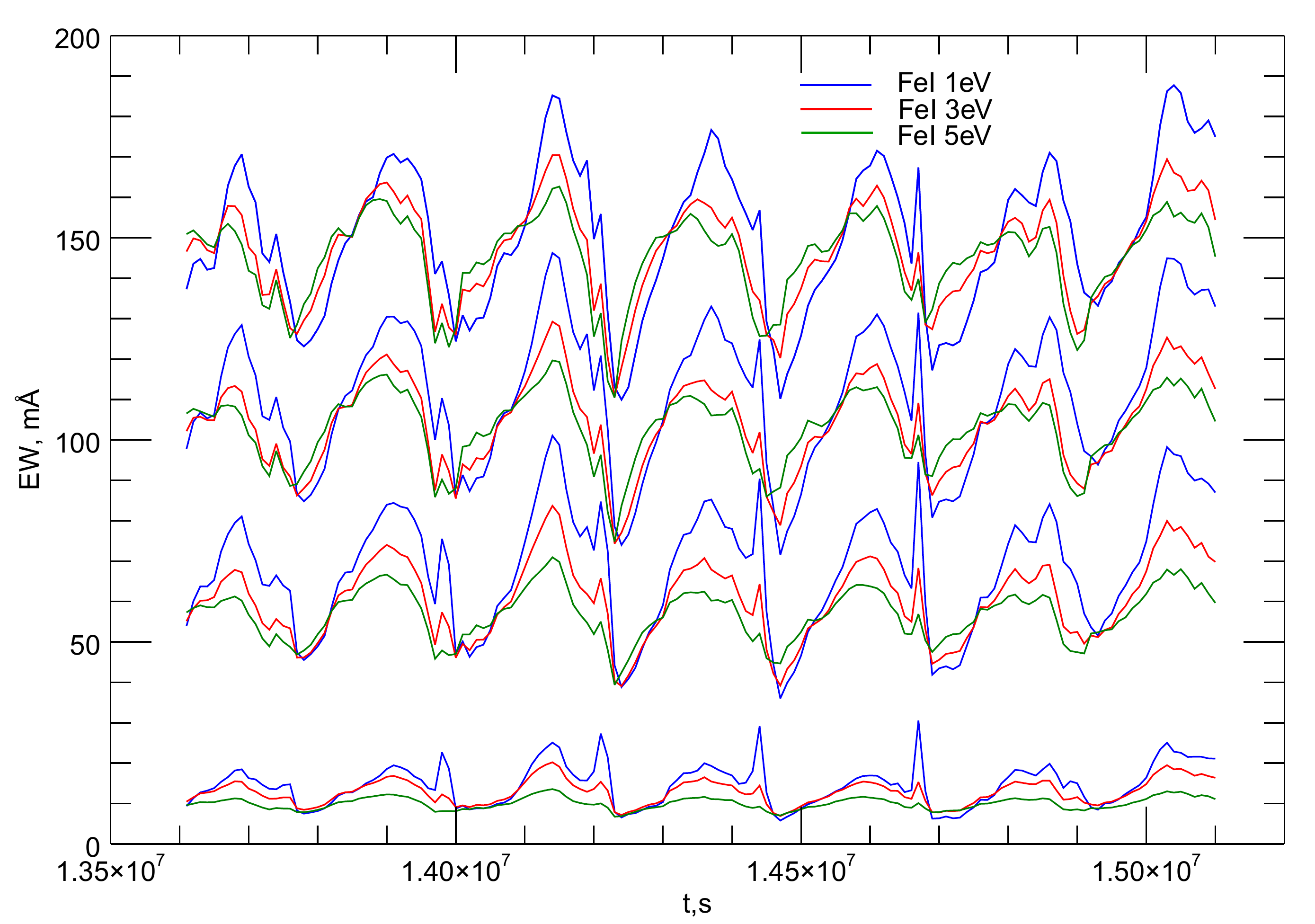}
\caption{Equivalent widths as a function of time for lines of \ion{Fe}{i}
  $\lambda5500$\,\AA\ with excitation energy $E_i=1,3,5$\,eV. The EWs are
  calculated by 
  replacing the 2D hydrodynamic velocity field by an isotropic,
  depth-independent Gaussian microturbulent velocity of
  $\xi_\mathrm{mic}=3$\kmos.}
\label{EWvmic3FeI}%
\end{figure}

For step (ii), the spectral synthesis was calculated using a
depth-independent isotropic Gaussian microturbulent velocity in the range
$\xi_\mathrm{mic}=0.5-6.0$\kmos\ to cover the range usually found in studies
of Cepheids \citep{1998AJ....115..605L,2002yCat..33920491A}. A plot with the
result of the spectral synthesis of the \ion{Fe}{i} $\lambda5500$\,\AA\ with
$\xi_\mathrm{mic}=3$\kmos\ is shown in Fig.~\ref{EWvmic3FeI}.  The increase of
the microturbulent velocity leads to an increase of the EW. Strong lines are
very sensitive to variations of the microturbulence.
 
\begin{figure}
\includegraphics[width=\hsize]{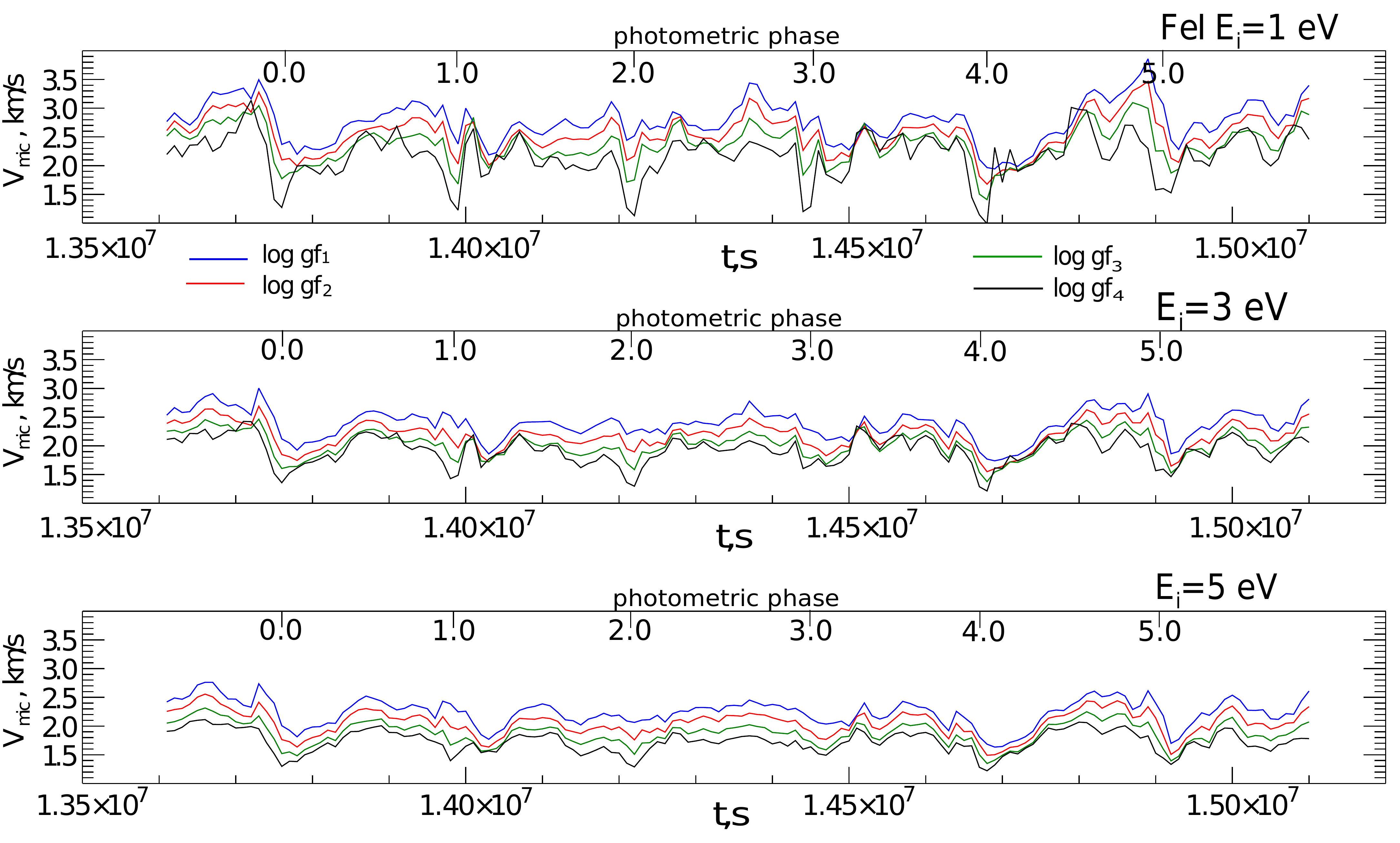}
\caption{Microturbulent velocity $V_\mathrm{mic}$ as a function of time for
  lines of \ion{Fe}{i} $\lambda5500$\,\AA\  with $E_i=1,3,5$\,eV.
  $\log\mathit{gf}_1$ (strongest line) $>\log\mathit{gf}_2 > \log\mathit{gf}_3
  >\log\mathit{gf}_4$ (weakest line).}
\label{vmicFeI}%
\end{figure}

\begin{figure}
\includegraphics[width=\hsize]{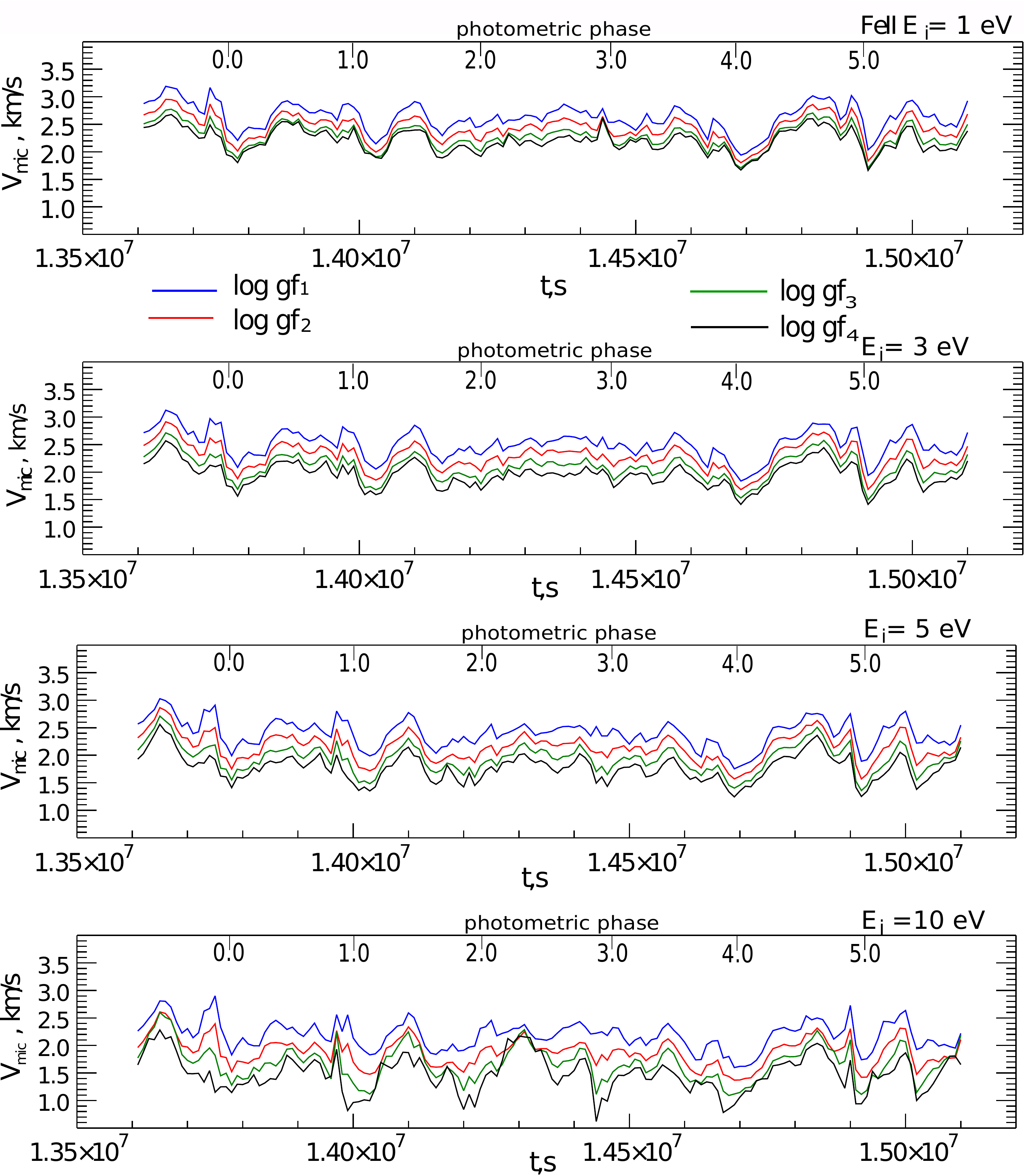}
\caption{Like Fig.~\ref{vmicFeI} but for line of \ion{Fe}{ii} 
$\lambda5500$\,\AA\ with $E_i=1,3,5,10$\,eV.}
\label{vmicFeII}%
\end{figure}

The calculation of $\xi_\mathrm{mic}$ by the requirement
$\mathrm{EW}(v_\mathrm{2D})=\mathrm{EW}(\xi_\mathrm{mic})$ was done with a
third order spline interpolation. The microturbulence as a function of time
for \ion{Fe}{i} and \ion{Fe}{ii} lines is shown in Figs.~\ref{vmicFeI} and~\ref{vmicFeII}. We note that the calculated synthetic spectra have
infinite resolution and signal-to-noise, which allows us to assign a
microturbulence even to the weakest lines of our sample. The microturbulence
exhibits a clear modulation with the pulsational phase with an amplitude of
$\approx 0.5 \ldots 1.0$\kmos\ and a certain degree of randomness due to
convection. Around photometric phase 0.9 before  maximum light
the microturbulent velocity curve exhibits distinct peaks. These peaks coincide with
the moment when the model is near its minimum radius.  
This is qualitatively  in agreement with the temporal behaviour of  turbulent velocities as measured by \cite{2017arXiv170700738B}  using an autocorrelation technique. 
One has to keep in mind that their turbulent velocities include contributions from  rotation as well as macro- and microturbulence.  The absolute value of the microturbulent velocity is $1.0 \ldots
4.0$\kmos, which is less than observed in short period Cepheids $\approx 1.5
\ldots5.0$\kmos\ \citep{2001yCat..33810032A}. The microturbulence slightly
decreases with increasing excitation potential, and it averaged
over all lines varies with pulsational cycle between 1.5 and 2.7\kmos.

There are several reasons why our model may fall somewhat short of the
observed values: (a) our procedure to derive the microturbulence does not
exactly mimic the observational procedure; (b) our selection of lines differs
noticeably from the set of lines used in the observations; (c) our model
harbours indeed a low microturbulence. In particular, limited spatial
resolution leads to a systematically low $\xi_\mathrm{mic}$
\citep{2013MSAIS..24...37S}. Nevertheless, we are content that at least
on the whole the microturbulence in the 2D model follows the evidence from
observations.

\subsection{Spikes in the EW(t) curves}

\begin{figure}
\includegraphics[width=\hsize]{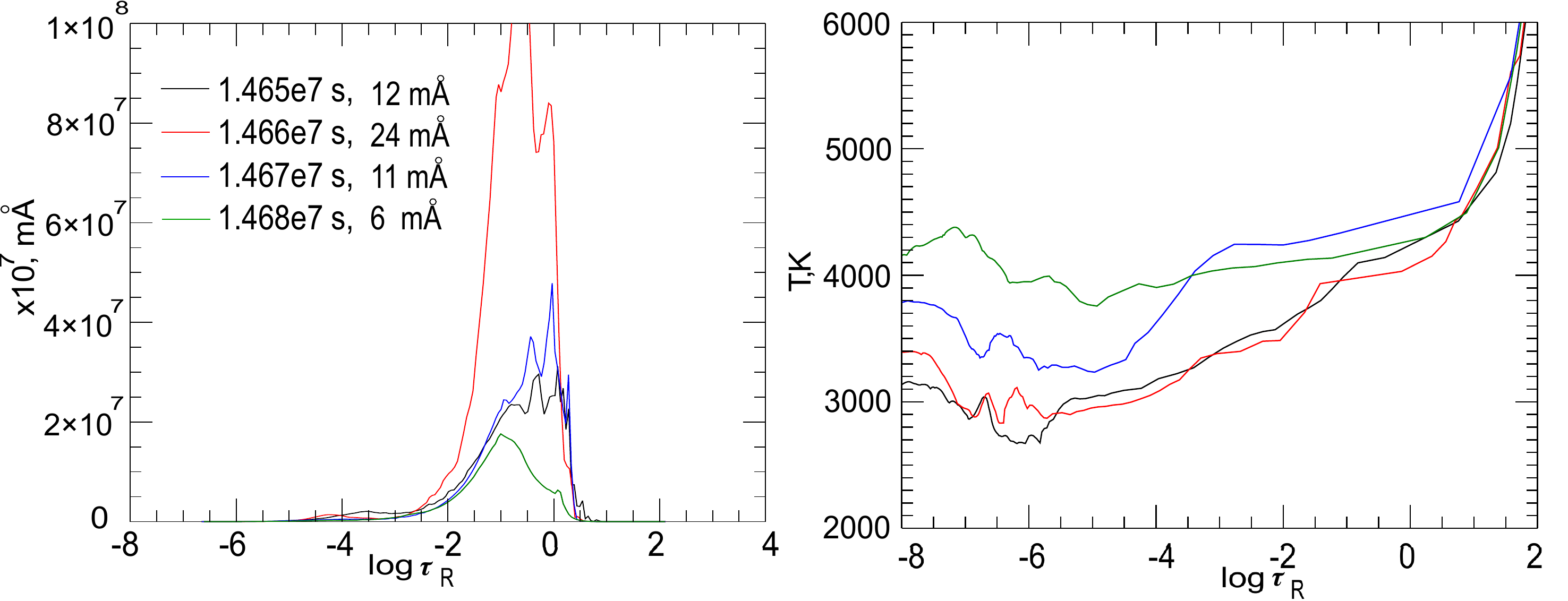}
\caption{Time dependence of the non-normalized line equivalent
  width contribution functions of the  weakest \ion{Fe}{i}
  $\lambda5500$\,\AA\ $E_i=1$\,eV line (left panel) and  the
  temperature profiles (right panel). The contribution functions and the temperature profiles  for each instance in
  time are depicted in corresponding colours.}
\label{spike}%
\end{figure}

Both results of the spectral synthesis coming from step (i) with the full 2D
model and step (ii) with the microturbulent velocity show that the EWs
depend on the pulsational phase. The amplitude of the variation of the EW is
larger for stronger lines. Occasionally, the EWs from steps (i) and (ii)
show sharp spikes. This behaviour is obviously not related to the 2D
velocity field, because the spikes still appear in step (ii) where the velocity
field is replaced by a constant microturbulent velocity. The change of the
thermal structure plays the key role here, which is illustrated in
Fig.~\ref{spike}. At time $t=1.466 \cdot 10^7$\,s, the equivalent width of the
\ion{Fe}{i} lines in Fig.~\ref{FeIew} has a sharp maximum because the temperature gradient is
steeper than at previous and subsequent times in the line formation region
with Rosseland optical depth $\tau_\mathrm{R}=10^{-2} \ldots 1$. The spikes
of the equivalent width of the \ion{Fe}{i} lines  show up during phases 
of maximum compression.   The EWs of the \ion{Fe}{ii} lines show 
additional  spikes  due to the deep location of the line formation  regions.
The impact of convection on  the \ion{Fe}{ii} line formation regions 
is  more than for similar regions of  \ion{Fe}{i}  lines.
Lines forming very deep might also reach into the region of the badly resolved
sub-photospheric temperature jump. Unfortunately, in this important region, where convection
starts, the numerical resolution is not quite sufficient.

\subsection{Line asymmetry}

\begin{figure}
\includegraphics[width=\hsize]{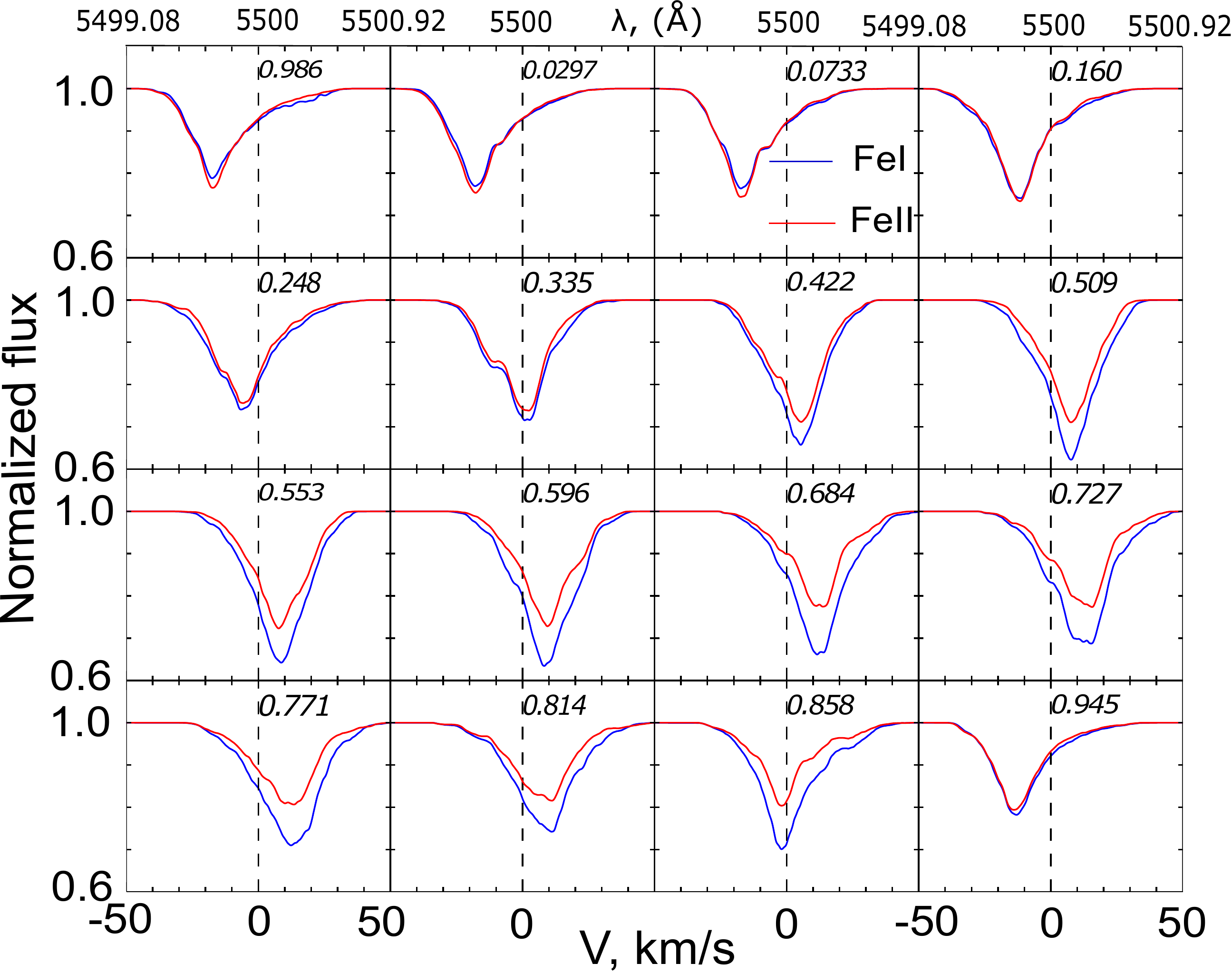}
\caption{\ion{Fe}{i} $E_\mathrm{i}=1$\,eV and \ion{Fe}{ii} $E_\mathrm{i}=1$\,eV
  line profiles for the strongest case as a function of photometric phase.
  Photometric phase  $\phi_\mathrm{p}$ zero coincides with maximum light. }
\label{lineprof}%
\end{figure}

\citet{1995ApJ...446..250S} devised a method to measure the asymmetry of
lines. They found that for classical Cepheids the observed line asymmetry is a
function of the pulsational phase and correlates with the position of the line
core.  The asymmetry of a spectral line profile was determined from the
difference and sum of the areas of the red and blue profile halves:
$(S_\mathrm{red} - S_\mathrm{blue})/S_\mathrm{tot}$. Two different methods
were used to determine the line core position:
\begin{enumerate}
\item fitting a Gaussian to the whole line profile,   
\item fitting a parabola to the line core below  70\,\% of the continuum level. 
\end{enumerate}
The results of the spectral syntheses show a complex multicomponent
structure of the line profiles. However, we found that 30\,\% is a reasonable
part of the line to limit systematic effects related to the complex line
profile on the determination of the position of the line core.
\ion{Fe}{i} $E_\mathrm{i}=1$\,eV and \ion{Fe}{ii} $E_\mathrm{i}=1$\,eV line
profiles for the strongest case as a function of the photometric phase are
shown in Fig.~\ref{lineprof}. The photometric phase zero coincides with maximum light of the light curve.  Synthetic line profiles for most phases have
a multicomponent structure.

\begin{figure}
\includegraphics[width=\hsize]{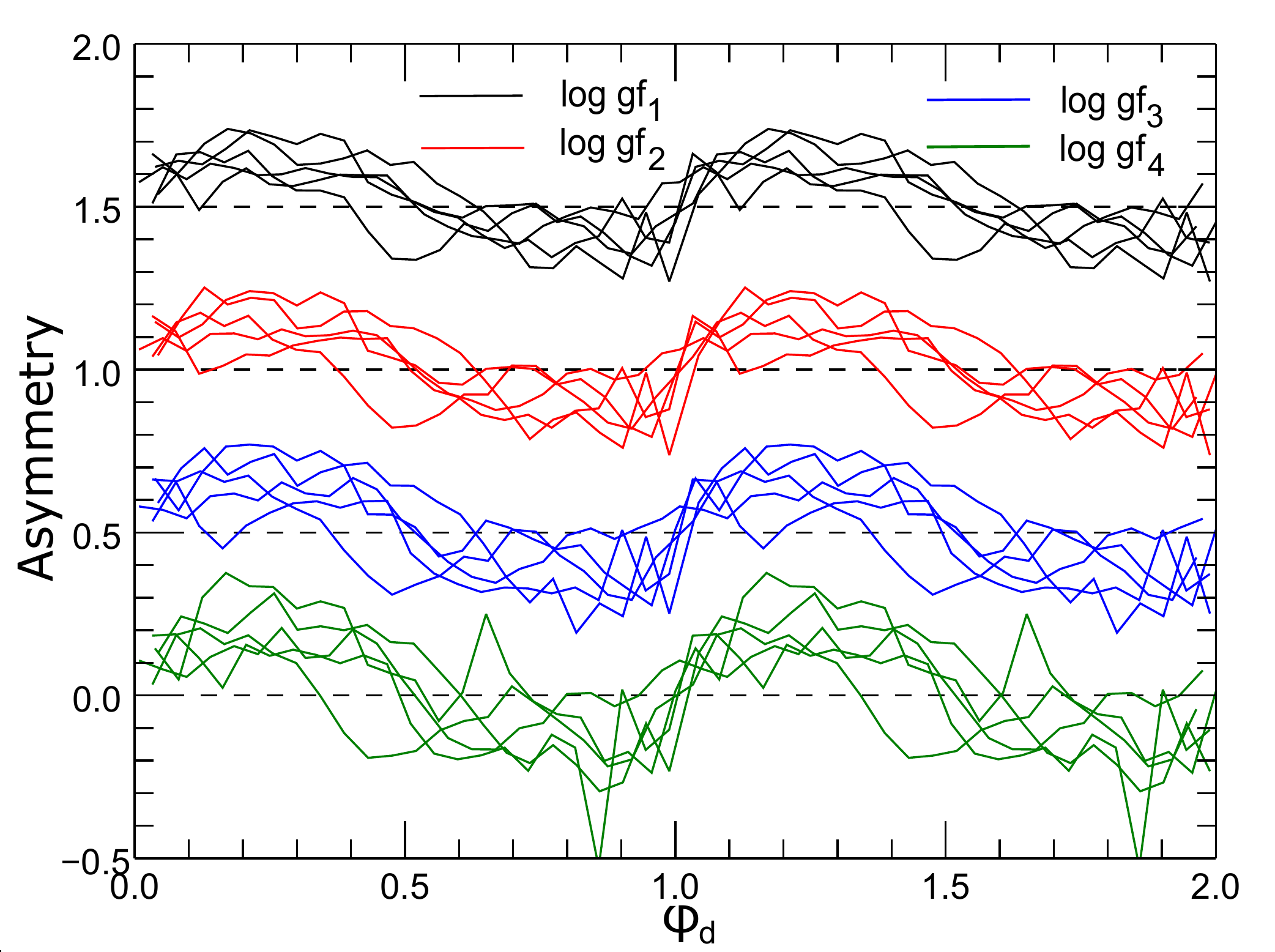}
\caption{Line asymmetries of the \ion{Fe}{i} $\lambda5500$\,\AA\ with
  $E_i=3$\,eV as a function of the dynamical phase $\phi_\mathrm{d}$. Line strength changes
  in order of oscillator strength with 
$\log\mathit{gf}_1>\log\mathit{gf}_2 > \log\mathit{gf}_3 >\log\mathit{gf}_4$.
  Dynamical phase zero coincides with the velocity reversal
  from contraction to expansion in the photosphere.  The line asymmetry  curves are shifted  by 0.5 dex relative to each other for clarity.}
\label{asssymFeI3ev}%
\end{figure}

Following \cite{1995ApJ...446..250S}, the line asymmetry was considered as a
function of the dynamical phase  $\phi_\mathrm{d}$ (see
Fig.~\ref{asssymFeI3ev}). The dynamical
phase zero coincides with the moment of the velocity reversal from  contraction
to expansion in the photosphere.  The amplitude of the line asymmetry
variation is $\approx 0.2 \ldots 0.3, $ which on average is larger than that found
in observational data. Sabbey et al. measured the amplitude of
variations of the line asymmetry to $\approx 0.1 \ldots 0.3 $. Nevertheless, the
time dependence of the line asymmetry has the same behaviour for the model and
observations \citep{1995ApJ...446..250S}. The 2D model provides only very
limited spatial statistics. We expect that the higher degree of smoothing when
averaging over the full stellar disc is reducing line asymmetries and
variations in line core position, largely removing the part that is stemming
from convection-related inhomogeneities. However, there is clear indication
that weak lines show systematically larger variations of the line asymmetry
independent of the method used to measure the line core position.

\subsection{Line-depth ratio and the effective temperature}

The ratio of depths of the spectral lines with high- and low-excitation
potentials is a very sensitive temperature indicator
\citep{1991PASP..103..439G,1994PASP..106.1248G}. This powerful method for
estimating the effective temperature was applied by
\citet{2006MNRAS.371..879K} and \citet{2001PASP..113..723G} to giant stars, and
by \citet{2000A&A...358..587K} and \citet{2007MNRAS.378..617K} to
supergiants. The calibration of \citet{2000A&A...358..587K} was used by
\citet{2005AJ....130.1880A}, \citet{2004AJ....128..343L}, and
\citet{2005AJ....129..433K} to determine the effective temperatures of
Cepheids over a range of periods from three to ten days.

\begin{figure}
\includegraphics[width=\hsize]{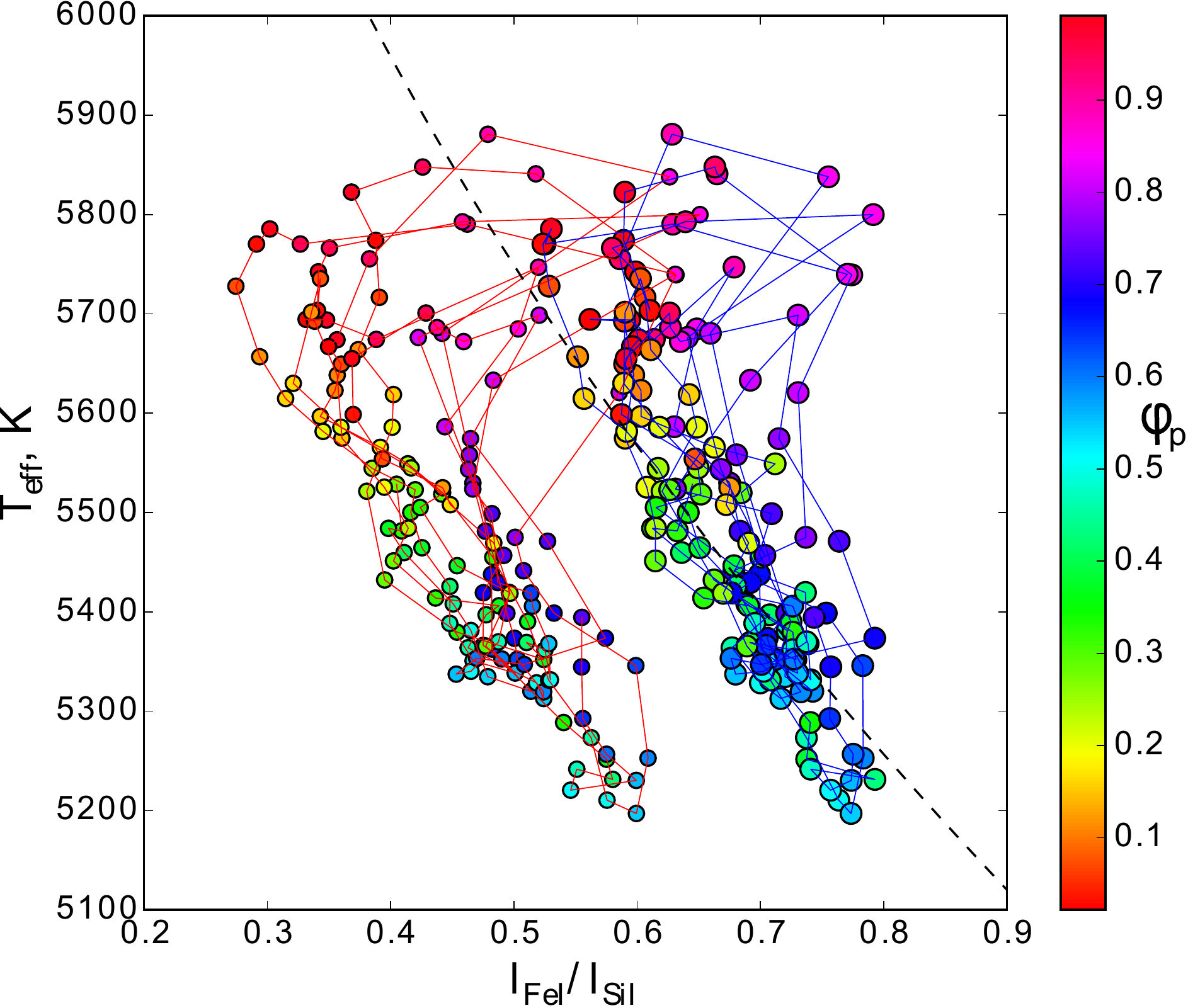}
\caption{Effective temperature as function of the \ion{Fe}{i}
  $\lambda6085.27$\,\AA\ and \ion{Si}{i} $\lambda6155.14$\,\AA\ line depth
  ratio for two different abundances. Big circles connected by a blue line
  mark a case with $+0.4$\,dex abundance increase for Si and Fe; small circles
  connected by a red line mark a case with $-0.4$\,dex. The colour of the
  circles indicates the photometric phase $\phi_\mathrm{p}$.  The dashed curve depicts the
  calibration of \cite{2007MNRAS.378..617K} for supergiants.}
\label{line_depthFeISiI}%
\end{figure}

\citet{2007MNRAS.378..617K} obtained around 130 different line pair relations
for estimating the effective temperatures of supergiants as a function of line
depth ratio. As one example, we investigated the often used pair \ion{Fe}{i}
$\lambda6085.27$\,\AA\ and \ion{Si}{i} $\lambda6155.14$\,\AA. The effective
temperature as a function of the \ion{Fe}{i} $\lambda6085.27$\,\AA\ to
\ion{Si}{i} $\lambda6155.14$\,\AA\ line depth ratio
$l_\mathrm{FeI}/l_\mathrm{SiI}$ in the 2D model is shown in
Fig. \ref{line_depthFeISiI} for two different solar-scaled abundances of Fe
and Si. To calculate the photometric phase $\phi_\mathrm{p}$ the radial velocity curve was
fitted by a harmonic function $A\cdot\sin(\omega t+ \phi_\mathrm{o}).$

The slope of the function
$T_\mathrm{eff}(l_\mathrm{\ion{Fe}{i}}/l_\mathrm{\ion{Si}{i}})$ is somewhat
steeper than in the case of supergiants (\cite{2007MNRAS.378..617K}).  It is
possible that for Cepheids with periods shorter than three\,days, dynamical
effects in the atmosphere are stronger. Again, cycle-to-cycle variations due
to the convective noise are clearly discernible. Interestingly, maximum
$T_\mathrm{eff}$ is compatible with range of values for  the line depth ratio, making
the overall relation not unique, and a small line depth ratio can hint towards
a low or high $T_\mathrm{eff}$. However, such an ambiguity can be easily
resolved if the phase of observation is roughly known. All in all we can
confirm that the method of line depth ratios works for our 2D model. However,
we see more fine-structure than a simple unique relation. Moreover, the
calibration for supergiants does not fit perfectly the data of our short
period Cepheid model.

\section{Interpretation of the "K-term"}

\begin{figure}
\includegraphics[width=\hsize]{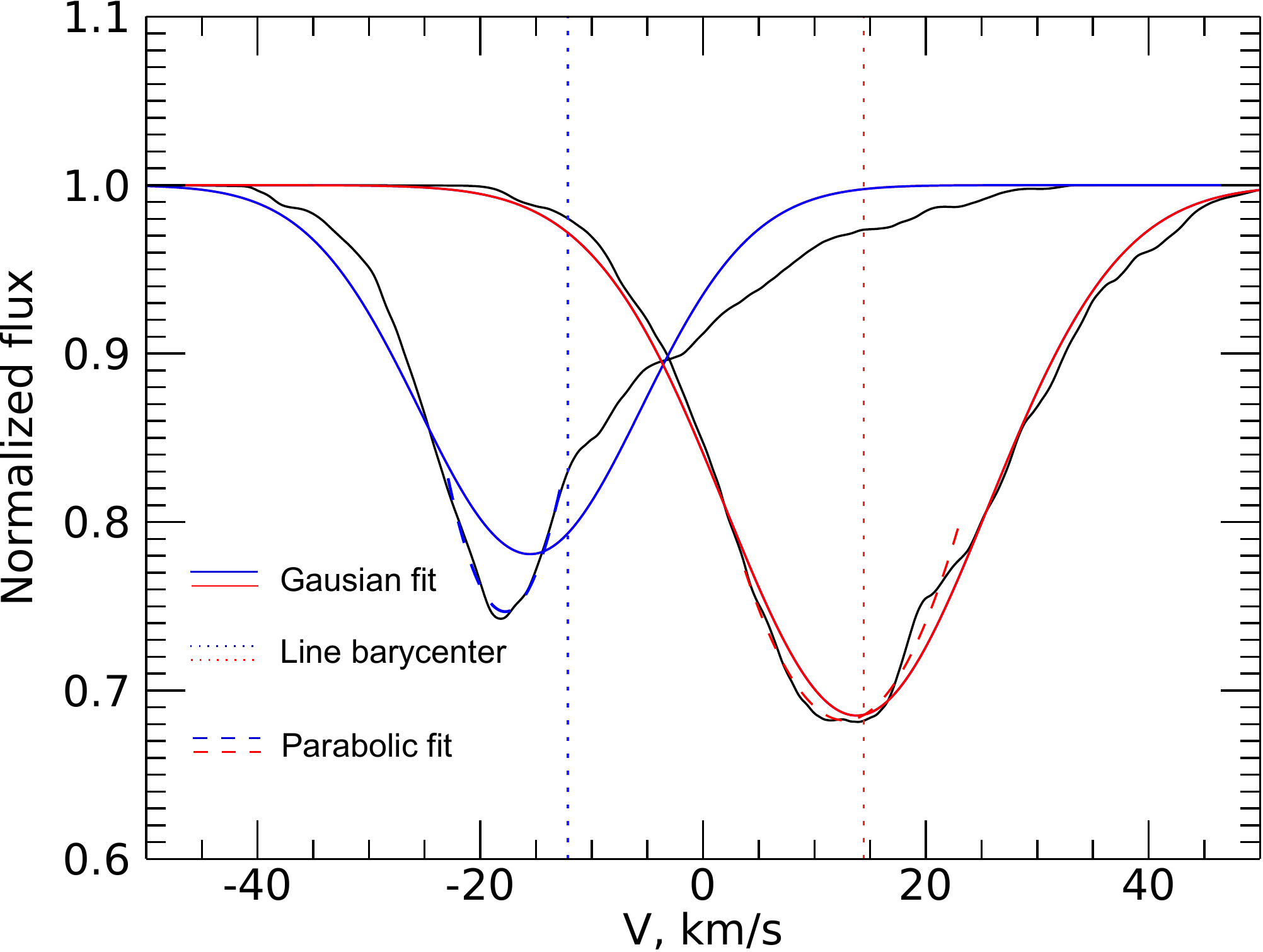}
\caption{The line profiles of \ion{Fe}{i} 1 $E_i=1$\,eV  for two extreme  cases (black) of  the minimum (blue) and maximum (red) of the spectroscopic radial velocity and  measured Doppler shifts of the the line-core position by using a parabolic fit to the core region,  by using a Gaussian fit to the whole profile  and  the centroid velocity. }
\label{lineshift_methods}%
\end{figure}

\begin{figure}
\includegraphics[width=\hsize]{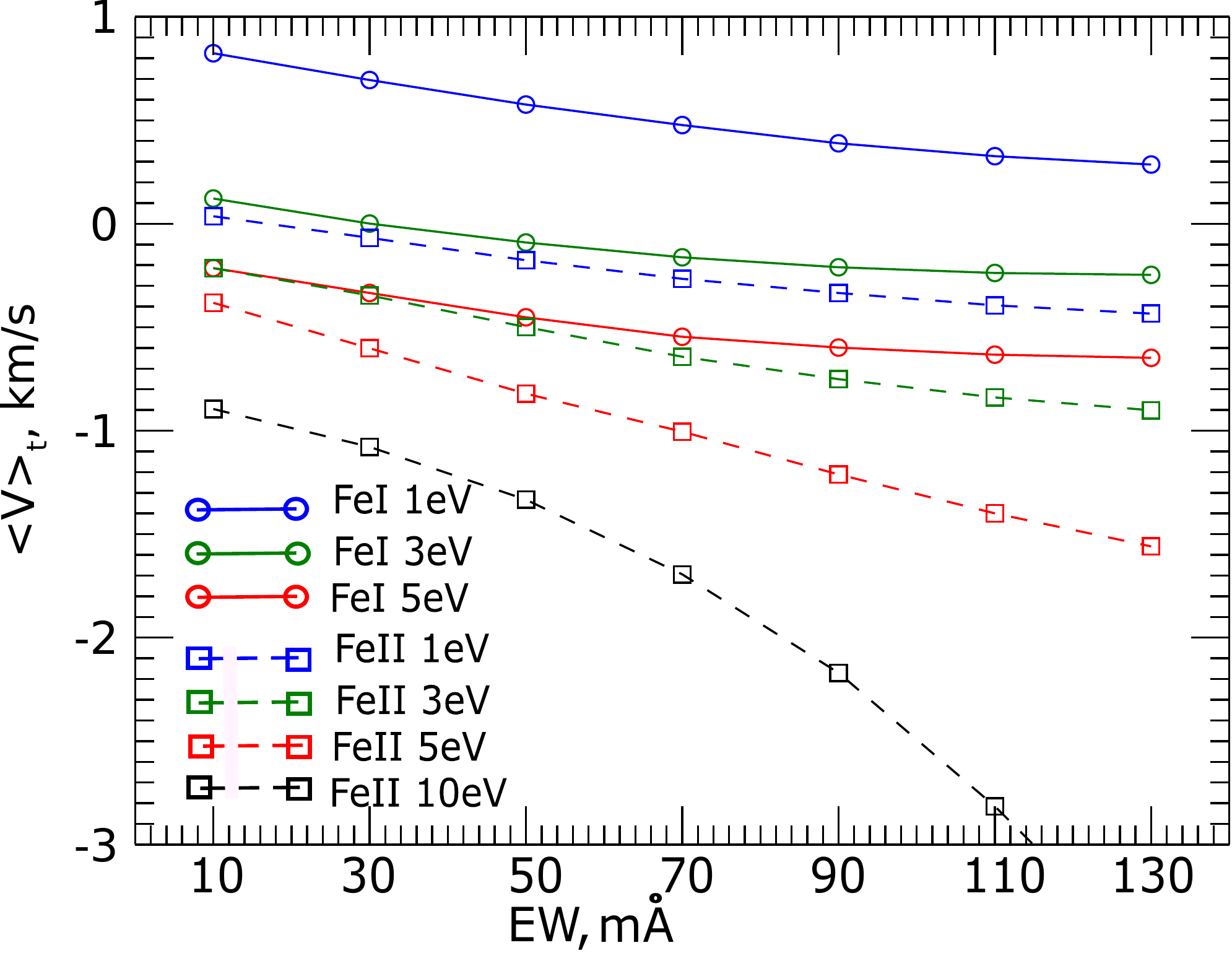}
\caption{Time-averaged radial velocities $\left\langle v \right
  \rangle_\mathrm{t}$ of \ion{Fe}{i} and \ion{Fe}{ii} lines as a function of
  their EW. The measurement of the Doppler shift is based on the  barycenter  of  the  spectral
line .}
\label{meanvelmasscenter}%
\end{figure}

As was emphasized by \cite{1995ApJ...446..250S}, different methods of
measuring the Doppler shift of absorption lines yield systematically different
results. For perfectly symmetric line profiles, measuring the line-core
position by using a parabolic fit to the core region (i), a  Gaussian
fit to the whole profile (ii), or the centroid velocity (iii), given by  the  first  moment of  the  spectral  line  profile, yields identical results.  The advantage of the third method is its independence  of  rotation and turbulent line broadening  \citep{1982A&A...109..258B, 2006A&A...453..309N}.   However, the absorption line
profiles of Cepheids are asymmetric. The three methods can thus lead to
different results, but fortunately -- as it turns out -- the qualitative
picture does not depend on the specific choice. The line profiles of 
\ion{Fe}{i} 1 $E_i=1$\,eV are shown  for two extreme  cases of  
the minimum and maximum  of the spectroscopic radial velocity and measured Doppler shifts with these methods in
the Fig.~\ref{lineshift_methods}. For the blue shifted  asymmetric line, the third method gives the smallest
Doppler shift.  The difference to a Gaussian fit  is $\approx5$ \kmos. For the red shifted case, the line profile
is less asymmetric and all methods give almost the same Doppler  velocity. From this experiment we expect  that
the moment method gives smaller blue-shifts, which is confirmed later (see Fig.~\ref{meanvelmasscenter}). We
remark that we neglect the  gravitational redshift in our discussion, which is $\lesssim 100$ \mos for Cepheids.

In the following the
spectroscopic radial velocity is based on measurement of the Doppler shift of
the absorption line using both these methods.

\begin{figure}
\includegraphics[width=\hsize]{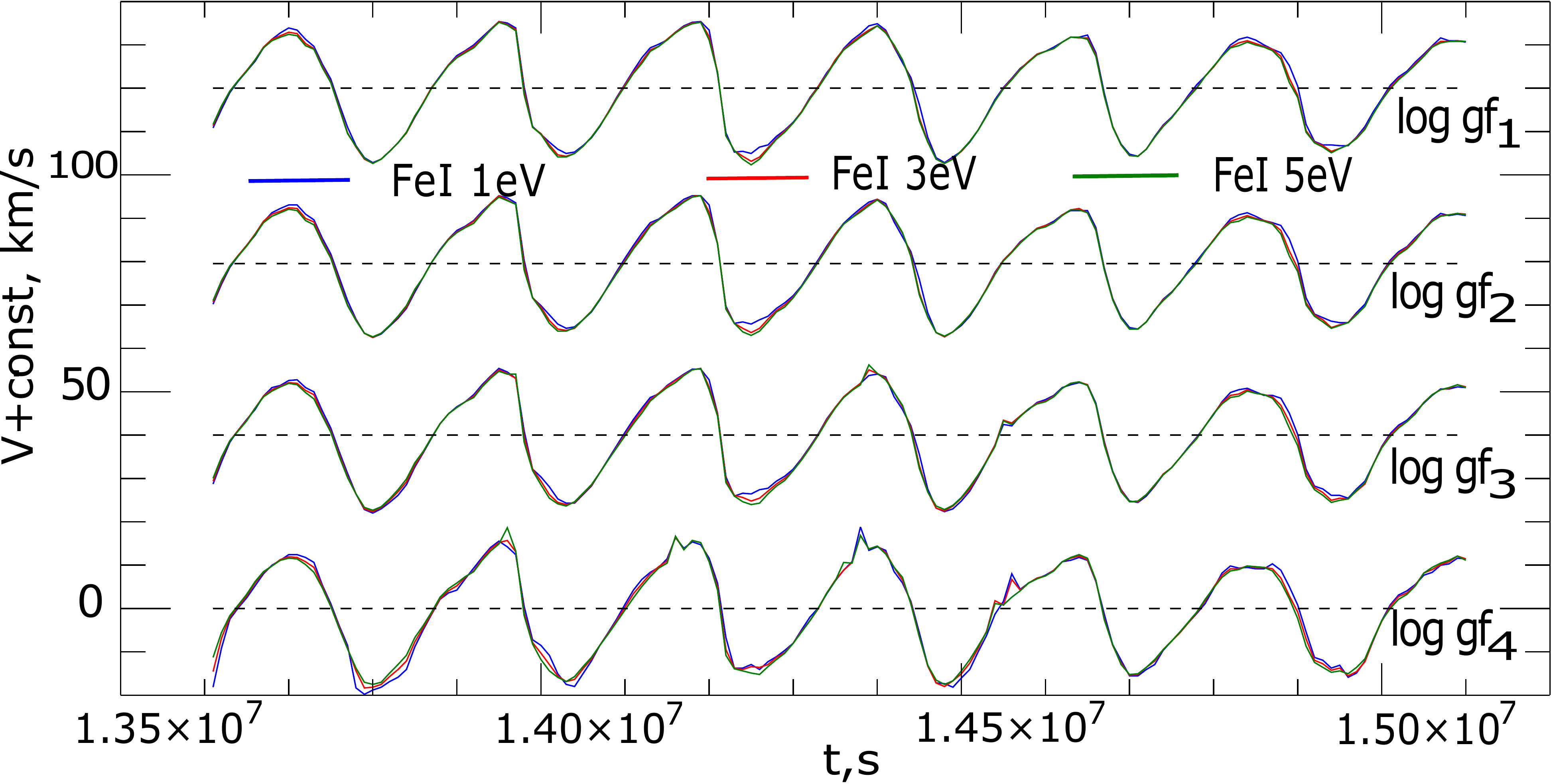}

\caption{Radial velocities as a function of time obtained by measuring the
  Doppler shift of \ion{Fe}{i} $\lambda5500$\,\AA\ $E_i=1$\,eV (blue), 3\,eV
  (red), 5\,eV (green) line cores using parabolic fits. Four different
  line strengths are shown whose EWs are ordered according to their oscillator
  strength $\log\mathit{gf}_1>\log\mathit{gf}_2 > \log\mathit{gf}_3 >\log\mathit{gf_4}$. The velocity curves are shifted  by 40 km/s 
  relative to each other for clarity.}
\label{RVspectroscopic}%
\end{figure}

\begin{figure}
\includegraphics[width=\hsize]{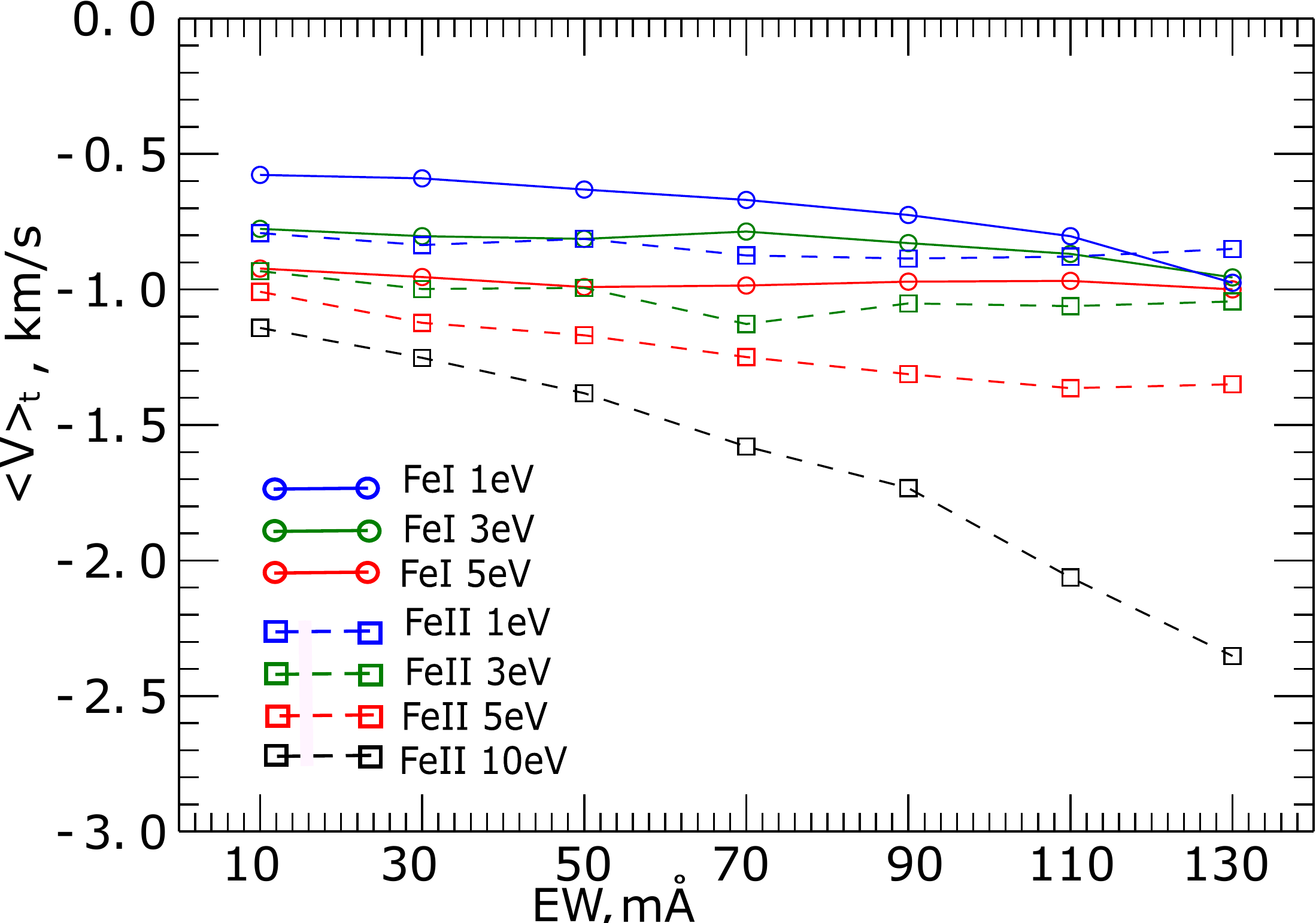}
\caption{Time-averaged radial velocities $\left\langle v \right
  \rangle_\mathrm{t}$ of \ion{Fe}{i} and \ion{Fe}{ii} lines as a function of
  their EW. The measurement of the Doppler shift is based on parabolic fits
  of the line core deeper than 70\,\% of the total line depth.}
\label{meanvelparab}%
\end{figure}

\begin{figure}
\includegraphics[width=\hsize]{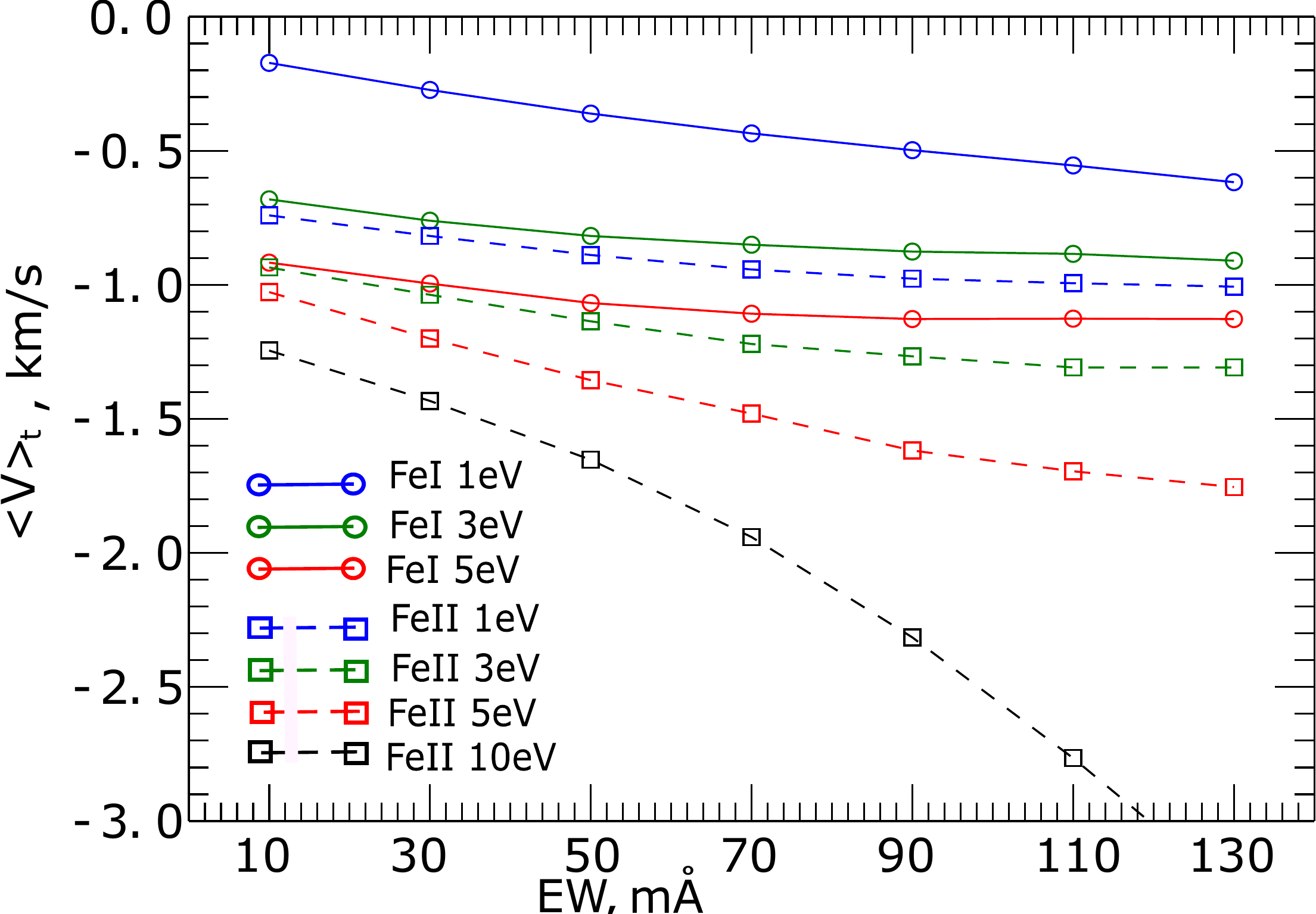}
\caption{Like Fig.~\ref{meanvelparab}, but measuring the Doppler shift
  by a Gaussian fit of the whole line profile.}
\label{meanvelgauss}%
\end{figure}

\begin{figure}
\includegraphics[width=\hsize]{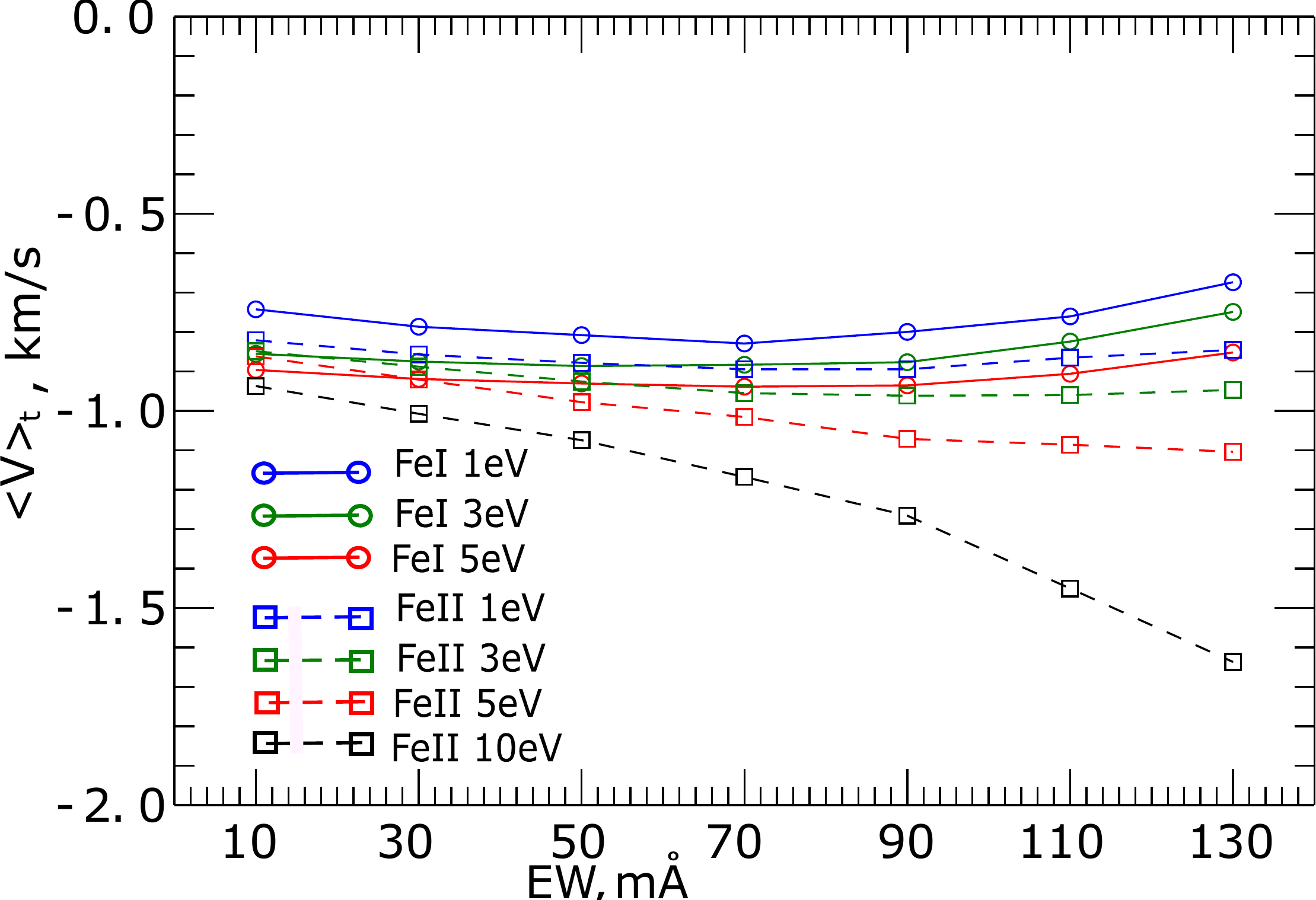}
\caption{Time-averaged radial velocities from Gaussian fitting when pulsations have
  not set in.}
\label{meanvel_init_gauss}
\end{figure}

The radial velocity does not show noticeable differences of the amplitude and
phase shift for different excitation potentials and oscillator strengths (see
Fig.~\ref{RVspectroscopic}). In contrast, the mean radial velocity $\langle v
\rangle_\mathrm{t}$ shown in Figs.~\ref{meanvelparab}, ~\ref{meanvel_init_gauss},  and~\ref{meanvelgauss},
which is the result of an averaging over six full cycles, exhibits
dependencies on excitation potential and line strength. Additionally, the path
conservation integral is not equal to zero, which is a basic assumptions in the
Baade-Wesselink method \citep{1987VA.....30..197G}:
\begin{equation}
\int_0^1 V_\mathrm{rad}(\phi) \ d\phi =  0,
\end{equation} 
where the integral is calculated over a complete pulsation cycle.  Path
conservation is used to derive the centre-of-mass velocity $V_\mathrm{\gamma}$
(or $\gamma$-velocity ) of a pulsating star from the above integral of the
radial velocity via the condition $\int_{ \ 0}^{1} (V_\mathrm{rad} -
V_\mathrm{\gamma}) \ d\phi = 0$.

Observational evidence for the presence of a residual systematic velocity has
existed for around 70 years
\citep{1947pezv.book.....P,1956MNRAS.116..453S,1974A&AS...15....1W,1995ApJ...446..250S}. The
kinematic behaviour of Cepheids
\citep{1991ApJ...378..708W,1994A&A...285..415P} shows a "K-term" , which is a
residual line-of-sight velocity of the order $-2$\kmos\ (toward us) between
the measured (spectroscopically) centre-of-mass velocity and the Galactic
rotational velocity derived from the kinematics of stellar
populations.  \cite{1995ApJ...446..250S} presumed that this residual velocity
is related to the varying depth of the photospheric line-forming region with
pulsational phase.   Based on very high quality the High Accuracy Radial velocity Planet Searcher (HARPS)
observations and careful
methodology, \cite{2009A&A...502..951N} showed that the $\gamma$-velocities are
linearly correlated with the line asymmetry~$A$ $V_\gamma = a_0 A +b_0$ and
intrinsic properties of Cepheids.  The residual $\gamma$-velocity (or K-term) averaged
over eight stars was measured to $b_0=1.0 \pm 0.2$\kmos.

\begin{figure}
\includegraphics[width=\hsize]{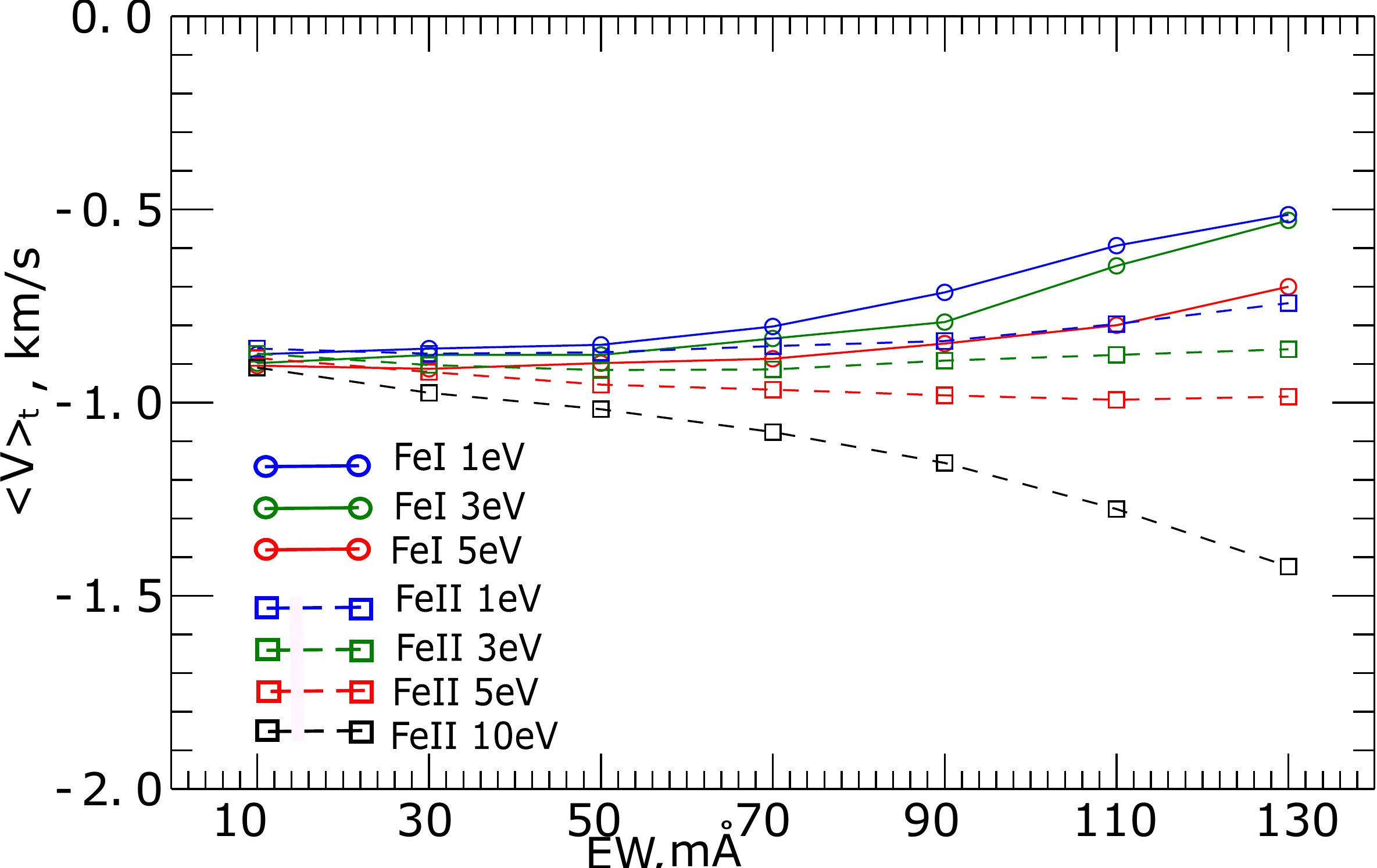}
\caption{Time-averaged radial velocities from line core fitting when pulsations have
  not set in.}
\label{meanvel_init_parab}%
\end{figure}

\begin{figure}
\includegraphics[width=\hsize]{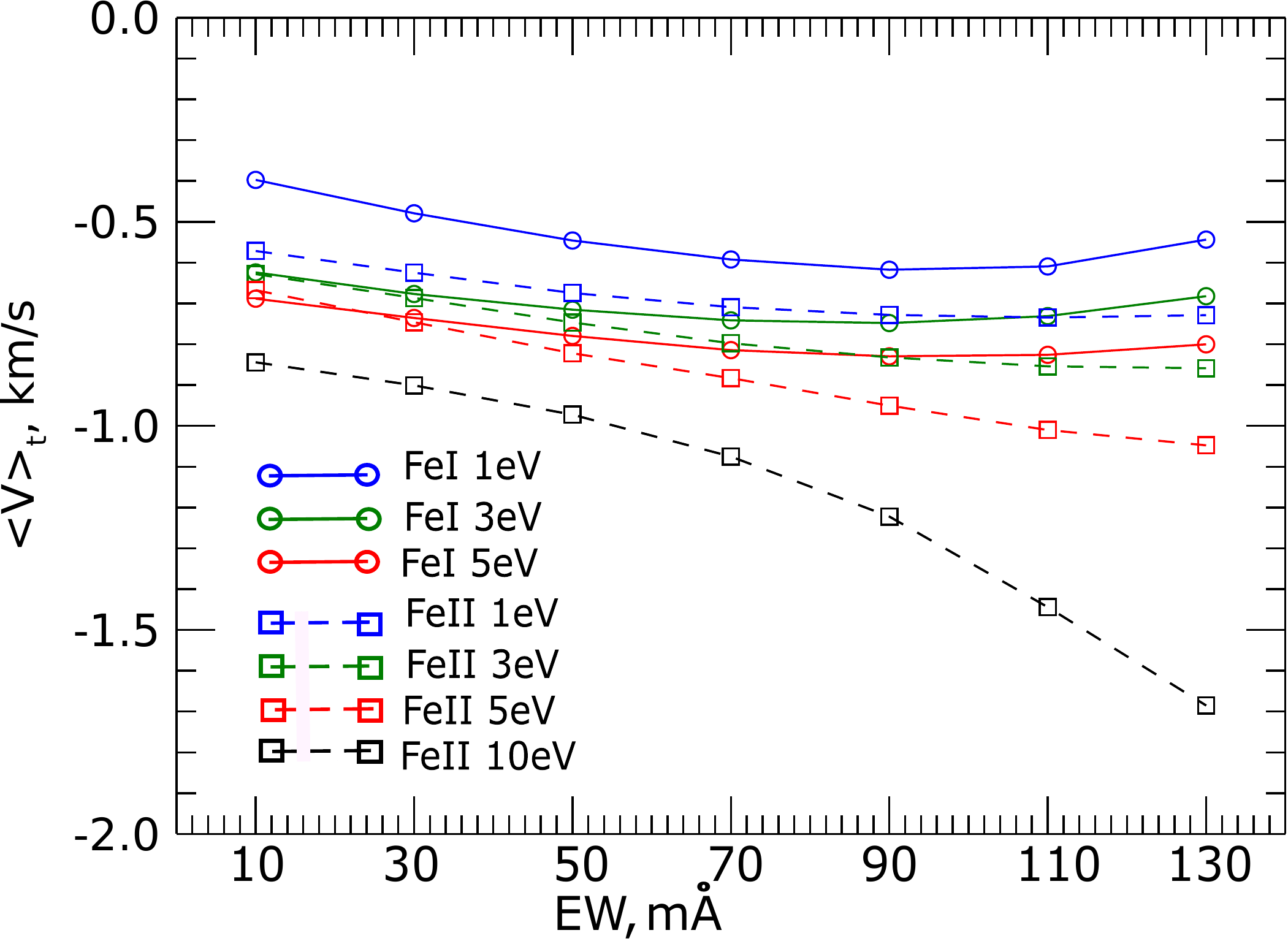}
\caption{Time-averaged barycenter radial velocities  when pulsations have
  not set in.}
\label{meanvel_init_masscenter}%
\end{figure}

The K-term can be affected by both pulsations and convection. To better
understand the contributions of the two processes, 2D model snapshots were
taken for an initial phase of the simulation run when pulsations had not set
in yet. Spectral syntheses of \ion{Fe}{i} ($E_i=1,3,5$\,eV) and \ion{Fe}{ii}
($E_i=1,3,5,10$\,eV) lines as in a case of pulsations were calculated for 100
2D snapshots between $(4.5-5.8)\times 10^6$ s. The Doppler shift of the lines
was measured using parabolic fits limited to residual fluxes of 30\,\% above
the line core, and Gaussian fits of the entire line profiles.  The
time-averaged radial velocities $\langle v \rangle_\mathrm{t} $ based on the
three methods are shown in the Figs.~\ref{meanvel_init_parab},~\ref{meanvel_init_gauss}, and~\ref{meanvel_init_masscenter}.  As was mentioned before, the method of fitting
does not change the qualitative picture and Figs.~\ref{meanvel_init_parab}, ~\ref{meanvel_init_gauss}, and~\ref{meanvel_init_masscenter} show a similar dependence of the mean
radial velocity on excitation potential, line strength, and ionization
degree. \ion{Fe}{ii} lines with high excitation potential are forming in
deeper layers of the atmosphere than in the case of the \ion{Fe}{i}.   This
indicates that the effect of convection is stronger than the effect of pulsations.

Comparing the mean velocities before and after the pulsations have set in
using both  Gaussian and parabolic fits and the first moment of the spectral line profile  (as shown in
Figs.~\ref{meanvel_init_gauss}, \ref{meanvelgauss}, \ref{meanvelparab},
\ref{meanvel_init_parab}, \ref{meanvel_init_masscenter}, and ~\ref{meanvelmasscenter})  led us to conclude that convection is the main
contributor to the residual velocity. The contribution of pulsations is not
essential, and only provides a spreading with line properties leaving the
averages largely intact. Therefore, we interpret the $\gamma$-velocity as the
Cepheid version of the convective blueshift present in all late-type stars
with convective outer envelopes.  The derived values of the residual
$\gamma$-velocity  varies between 1.0 and 0.5 \kmos dependent on method.  They are  close to the observed K-term for Cepheids \citep{2009A&A...502..951N}.  The dependence of  the centroid velocity on the line asymmetry in the first moment method leads to a smaller residual velocity of 0.5 \kmos (see Fig.~\ref{lineshift_methods}). 
A parabolic fit to the core region  gives $\approx1$ \kmos of the residual $\gamma$-velocity. The method  is  closer to the
cross-correlation method applied  for radial velocity measurements with the standard G2  mask with the HARPS spectrograph, because 
the widths of the "lines" of the mask  is 3 \kmos. This is much narrower than the width of lines in a Cepheid, and effectively leads to a determination of the line position  given by the line core.  The
period of the 2D model is 2.8\,days whereas typically Cepheids have longer
periods and lower surface gravities.   Spectroscopic observations 
of \cite{2006A&A...453..309N} indicate that for longer periods and lower $\log g$ 
the more vigorous convective motions produce larger residual $\gamma$-velocities.

\section{Projection factor obtained from the 2D model}

The projection factor~$p$ (also p-factor) is a key quantity in the
Baade-Wesselink method for determining the distances of Cepheids. The method
relates the spectroscopically measured radial velocities of a Cepheid to
interferometrically measured angular-diameters
\citep{1994ApJ...432..367S,2004A&A...416..941K}, or classically to their
  brightness, by using a calibrated surface-brightness-colour relation 
  \citep{1976MNRAS.174..489B,1997A&A...320..799F,1989ApJ...342..467G,2011A&A...534A..94S}. The
  method assumes the radius is defined at a specific location in the stellar
  atmosphere, for example, at an optical depth $\tau=2/3$. The change of the radius
  is $\Delta R =\int v_\mathrm{puls}\,dt$, where the pulsation velocity
  $v_\mathrm{puls}$ is the velocity of optical layers corresponding to the
  optical depth $\tau=2/3$.  The projection factor $p$ converts the
  spectroscopic radial velocity to a pulsational velocity: $v_\mathrm{puls}=p
  \cdot v_\mathrm{rad}$. Spectroscopic radial velocities include effects
  related to two kinds of integration over the stellar surface layers:
  horizontally, the effects of limb-darkening, and vertically, the effects of velocity gradients in the
  line-forming region.

The impact of limb-darkening and atmospheric expansion on the p-factor was
first studied by \cite{1926ics..book.....E} and
\cite{1928MNRAS..88..548C}. \cite{1972ApJ...174...57P}, using a model
atmosphere with uniform expansion, numerically determined the p-factor to be
between 1.31 and 1.34, depending on the width of a line.  Many attempts were
made to estimate and calibrate the p-factor from high-resolution spectroscopic
observations \citep{2004A&A...428..131N,2007A&A...471..661N,
  2009A&A...502..951N,2012A&A...541A.134N,
  2013A&A...553A.112N,2015A&A...576A..64B,
  2016A&A...587A.117B,2017A&A...597A..73N}.  \cite{2007A&A...471..661N} 
identified three effects affecting the projection factor: the geometrical
effect, the velocity gradient within the atmosphere, and the relative motion
of a layer of given optical depth with respect to a corresponding fixed Lagrangian mass
element.

\begin{figure}
\includegraphics[width=\hsize]{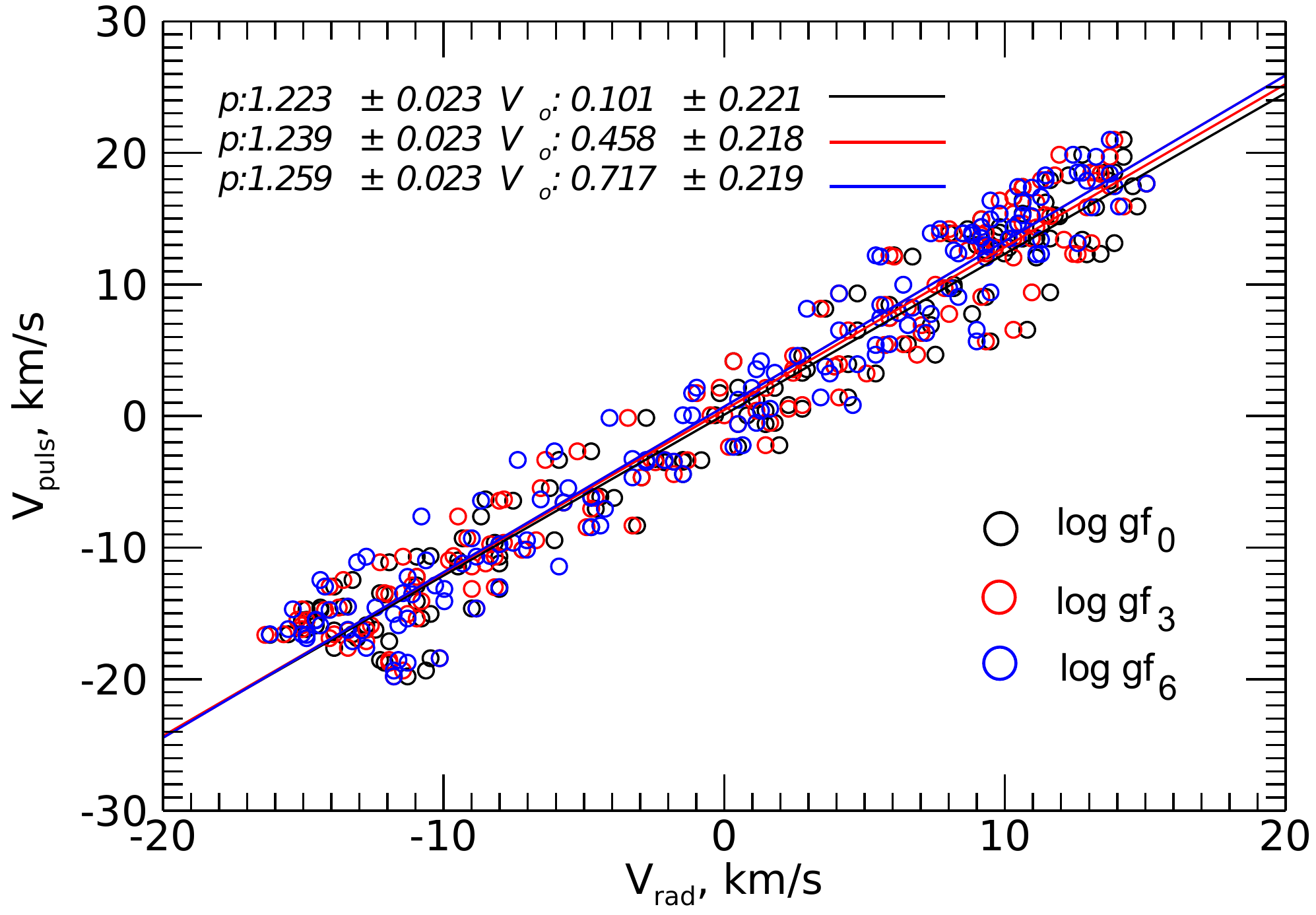}
\caption{Pulsational velocity as a function of the radial velocity,
  estimated by Gaussian fits of the whole \ion{Fe}{i} $E_i=1$ eV line
  profiles. Colours show different line strengths: the
  strongest line is depicted in black, the weakest in blue. The slope of the
  linear regression is the p-factor.}
\label{vpuls_vrad}%
\end{figure}

Our Cepheid model provides information about the pulsating velocity $
v_\mathrm{puls}(\tau_\mathrm{R}=2/3)$. The photospheric radius
$R_\mathrm{ph}(\tau_\mathrm{R}=2/3)$ was calculated with a cubic spline
interpolation using data of the mean 1D model. The pulsating velocity is
$v_\mathrm{puls}={dR_\mathrm{ph}}/{dt}$, which was calculated for each
instance in time over six full periods.  The projection factor is the slope of
the curve $v_\mathrm{puls}=p \cdot v_\mathrm{rad} + v_\mathrm{o} $ , which is
shown in Fig.~\ref{vpuls_vrad} for the line \ion{Fe}{i} $E_i=1$\,eV and three
different line strengths. We implemented the fitting to deal with the
substantial amount of convective noise present in the model data.
Depending on line strength the p-factor varies between 1.23 and 1.27,
which agrees with estimates from the literature \citep{2009A&A...502..951N}.  Additionally, within the
given noise level, Fig.~\ref{vpuls_vrad} shows the absence of a hysteresis-like
behaviour between the pulsational and radial velocities as was found when
investigating the relation between effective temperature and line depth ratio
(Fig.~\ref{line_depthFeISiI}).

\begin{figure}
\includegraphics[width=\hsize]{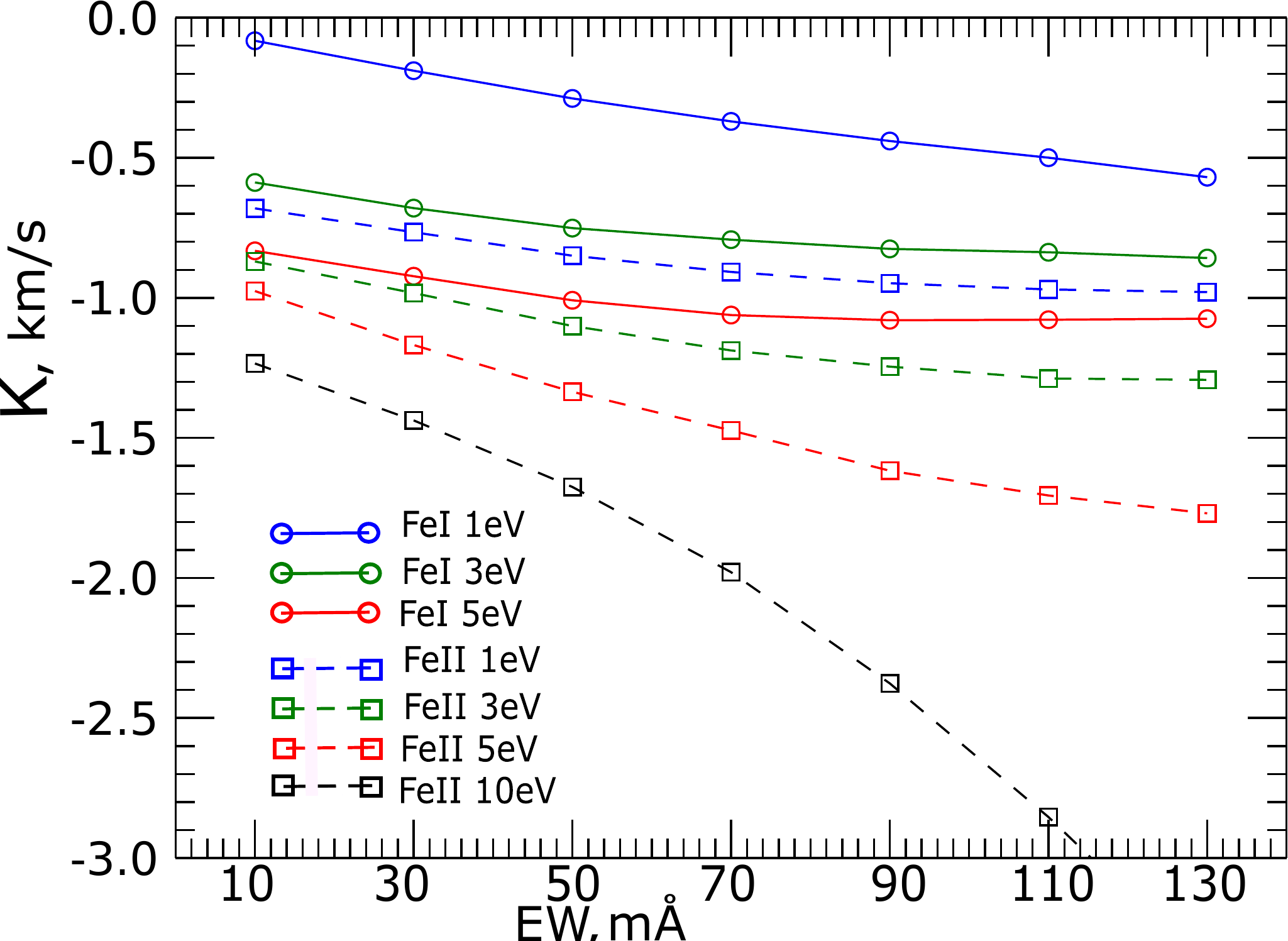}
\caption{Residual velocity or K-term derived by linear fits of the curve
  $v_\mathrm{puls}=p \cdot v_\mathrm{rad} + v_\mathrm{o} $ and
  $K=-v_\mathrm{o}/p$ for different line strengths and excitation
  potentials. The spectroscopic radial velocity is measured by Gaussian
  fitting.}
\label{K-term}%
\end{figure}

The fitting parameter $v_\mathrm{o}$ is related to the K-term. The presence of residual $\gamma$-velocity shifts the curve $v_\mathrm{puls}=p \cdot
v_\mathrm{rad} + v_\mathrm{o} $ to negative radial velocities along the
horizontal axis so that $K=-v_\mathrm{o}/p$.  The derived K-term from the
linear fits as a function of line strength is shown in Fig.~\ref{K-term}.

\begin{figure}
\includegraphics[width=\hsize]{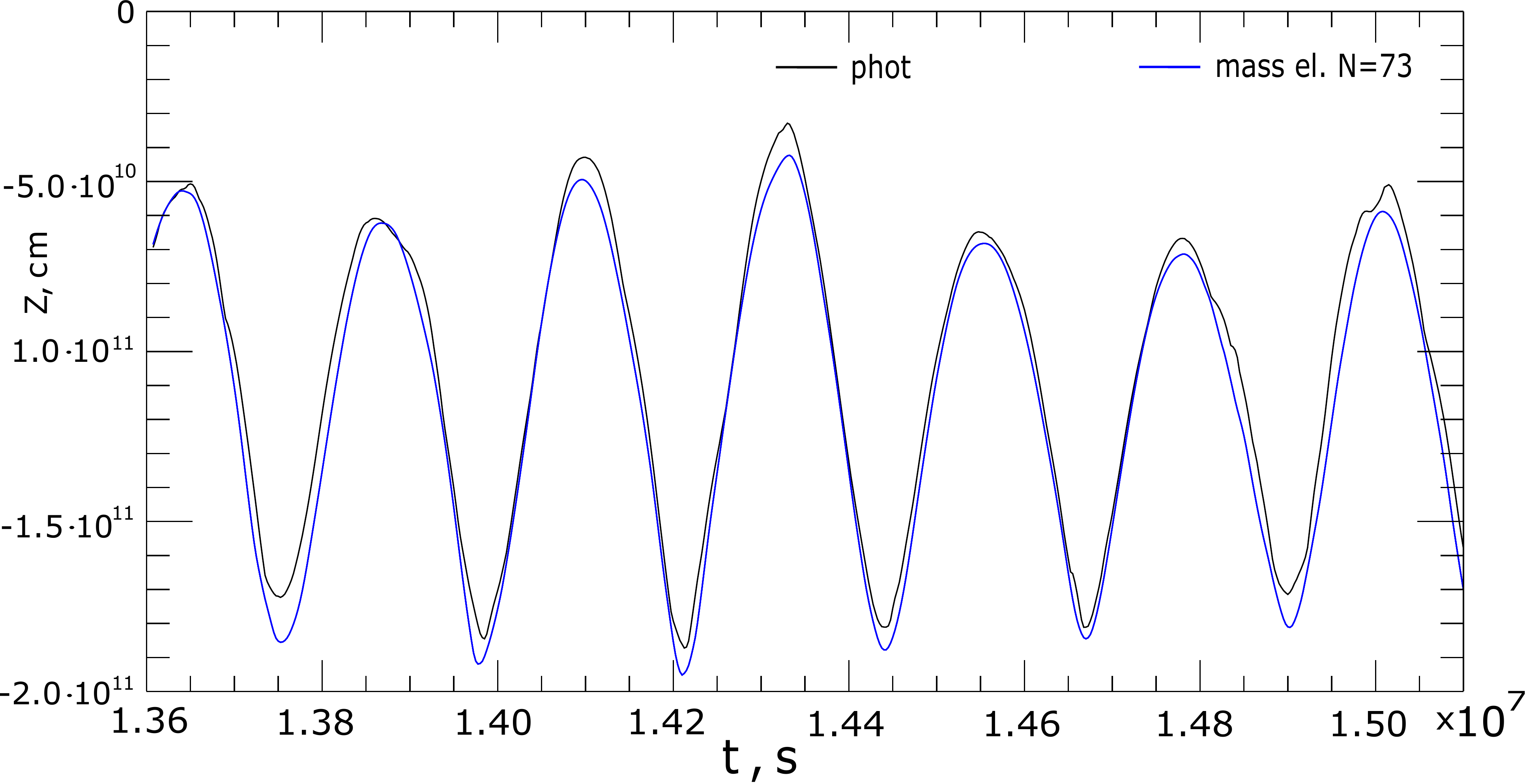}
\caption{Vertical geometric coordinate of a layer in the photosphere with
  $\tau_\mathrm{R}=2/3$, and the corresponding pseudo-Lagrangian mass element
  as a function of time.}
\label{lagrz}%
\end{figure}

\begin{figure}
\includegraphics[width=\hsize]{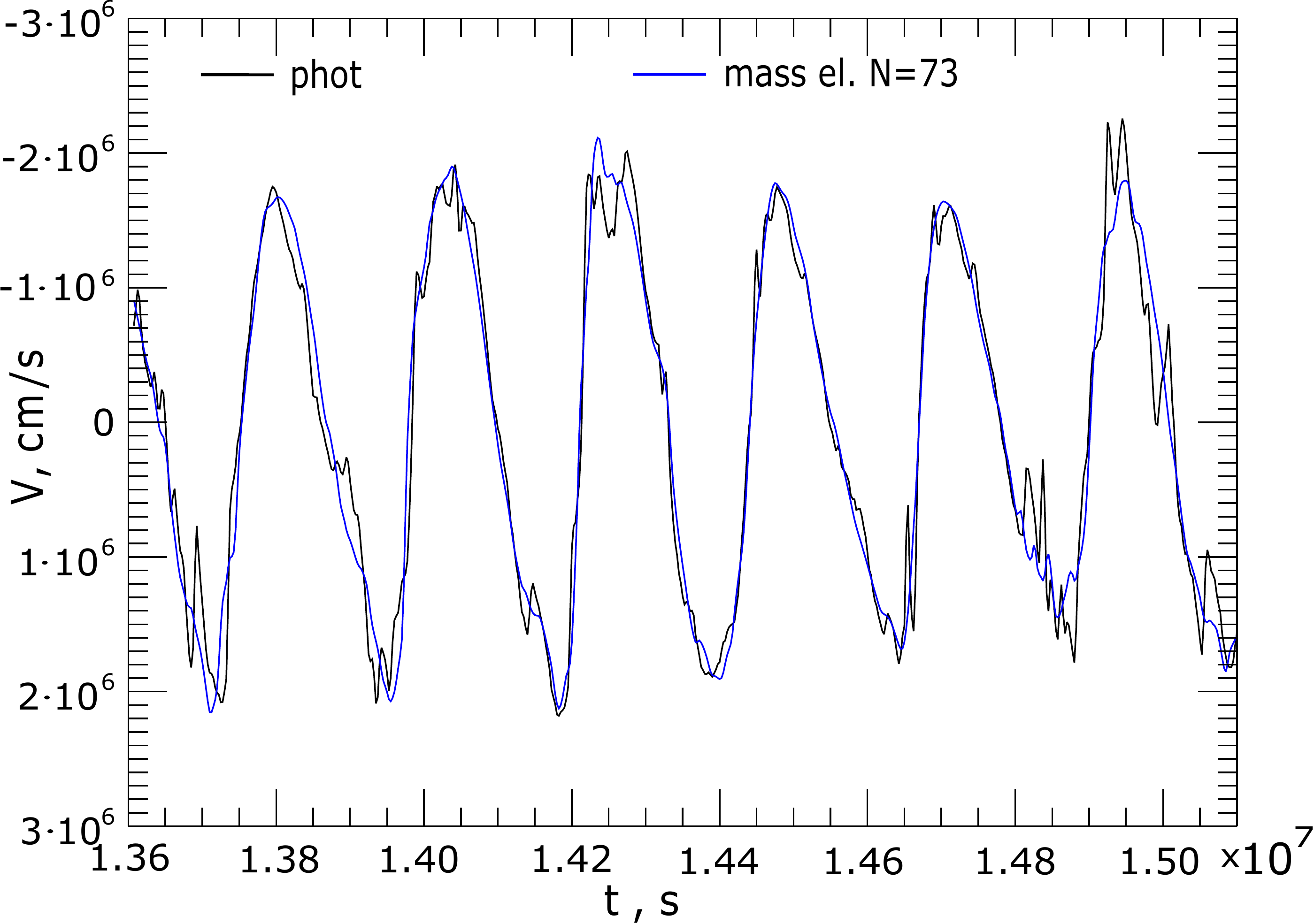}
\caption{Pulsational velocities of a layer in the photosphere with
  $\tau_\mathrm{R}=2/3$, and of the corresponding pseudo-Lagrangian mass
  element as a function of time.}
\label{lagrV}%
\end{figure}

From the physical point of view, the pulsational velocity is related to the
motion of mass elements in the atmosphere during the pulsations. One can
estimate the pulsational velocity using the velocities of mass elements close
to the photosphere, and thereby calculate the projection factor.  The
variation of the geometric vertical coordinate of a layer in the photosphere
with $\tau_\mathrm{R}=2/3$ and a Lagrangian mass element is shown in
Fig.~\ref{lagrz}. The pulsational velocity of these layers is shown in
Fig.~\ref{lagrV}. The pseudo-Lagrangian velocity of the mass element is almost
equal to the velocity of the optical depth surface $\tau_\mathrm{R}=2/3$ in
the photosphere. That means that in our model we find the same projection
factor for the pseudo-Lagrangian velocity, or the velocity of a layer of given
optical depth  (in the photosphere).

\section{Conclusions and final remarks}

A pilot study of spectroscopic properties of a multi-dimensional Cepheid
model was conducted. We arrive at the following conclusions:
\begin{enumerate}
\item The 2D model shows self-excited pulsations due to the $\kappa$-mechanism
  with a 2.8-day period. It reproduces the observed relation between the
  brightness and radial velocity amplitudes of pulsating stars
  \citep{1995A&A...293...87K}.
\item We find a net mean shift of the spectrum 
of the model Cepheid by $-0.5\ldots-1$ \kmos\
  which is a significant fraction of the observed residual line-of-sight
  velocity seen in the Cepheid population.
\item  We interpret this shift as the
  convective blueshift familiar from non-pulsating late-type stars.
  It is possible that multi-dimensional models with lower surface gravity
  (more similar to typical Cepheids) give higher convective and, eventually,
  higher residual velocities bringing the models closer to observations. This
  needs to be investigated further.
  
\item The microturbulent velocity, derived for individual iron lines, shows a
  modulation with pulsational phase and depends on the line properties. When
  averaged over all lines, it varies between 1.5 and 2.7\kmos\ over the
  pulsational cycle. This is somewhat lower than the velocity obtained in classical
  analyses with 1D hydrostatic atmosphere models
  \citep{2001yCat..33810032A}. Reasons for this shortcoming may be the
  limited spatial resolution of the model, or the different methodology used
  in the determination of the microturbulent velocity. However, around photometric phase 0.9 before  maximum light, the microturbulent velocity curve exhibits distinct peaks. This is qualitatively  in agreement with the temporal behaviour of    turbulent velocities  in  Cepheids as measured by  \cite{2017arXiv170700738B}  using an autocorrelation technique.
\item Line asymmetries show a behaviour that is qualitatively similar to
  observations. However, limited statistics makes it difficult to perform
  more quantitative comparisons. 
\item The projection factor derived from the 2D model lies between 1.23 and
  1.27, and agrees with observations \citep{2009A&A...502..951N}.
\end{enumerate}
Taken together we find a reasonable correspondence between the model and
observed properties of Cepheid variables. This is far from self-evident
considering the computational problems one faces when trying to simulate a
Cepheid envelope exhibiting self-excited pulsations together with violent
atmospheric dynamics. The 2D geometry together with detailed spectral
syntheses sets our model apart from most previous modelling works. It allowed
us to study the impact of convective inhomogeneities, which turned out to be
important for some spectroscopic properties -- here in particular for the
K-term -- and,  as we will present in a follow-up paper, for the
determination of elemental abundances. At least in the minds of the authors,
this leads to a shift in the way we should look at Cepheid variables. While
their defining property of being pulsating giant stars following a
period-luminosity relation is still perhaps their most important single
characteristic, now convection comes into focus. Its importance has always
been masked by the giant stars' prominent pulsations.

\begin{acknowledgements}
VV would like to thank Dr. Anish Amarsi for valuable comments on the draft.
HGL and BL  acknowledge financial support by the Sonderforschungsbereich SFB\,881 ``The
Milky Way System'' (subprojects A4, A5) of the German Research Foundation (DFG). 
The radiation-hydrodynamics simulations were performed at the
P{\^o}le Scientifique de Mod{\'e}lisation Num{\'e}rique (PSMN)
at the {\'E}cole Normale Sup{\'e}rieure (ENS) in Lyon.
\end{acknowledgements}


\bibliographystyle{aa} 
\bibliography{bibliography.bib}

\end{document}